\documentclass[12pt]{article}
\usepackage{amsmath,amssymb,latexsym,axodraw,cite}
\usepackage[]{hyperref}
\input xy
\xyoption{all}

\def\sqr#1#2{{\vcenter{\vbox{\hrule height.#2pt
            \hbox{\vrule width.#2pt height#1pt \kern#1pt
                  \vrule width.#2pt}\hrule height.#2pt}}}}

\def\square
 {\mathop{\mathchoice{\sqr{12}{15}}{\sqr{9}{12}}
{\sqr{6.3}{9}}{\sqr{4.5}{9}}}}

\begin{document}

\vspace*{0.5in}

\begin{center}

{\large\bf Decomposition in diverse dimensions}

\vspace*{0.2in}

Eric Sharpe \\
Department of Physics \\
Robeson Hall, 0435 \\
850 Drillfield Drive \\
Blacksburg, VA  24061
\\
{\tt ersharpe@vt.edu}

$\,$

\end{center}

This paper discusses the relationships between gauge theories defined by gauge
groups with finite trivially-acting centers, and theories with restrictions on
nonperturbative sectors, in two and four dimensions.
In two dimensions, these notions seem to coincide.  
Generalizing old results on orbifolds and abelian
gauge theories, we propose a decomposition of two-dimensional
nonabelian gauge theories with center-invariant matter into disjoint sums
of theories with rotating discrete theta angles; for example, schematically,
$SU(2) = SO(3)_+ + SO(3)_-$.  We verify that decomposition directly in
pure nonsupersymmetric two-dimensional Yang-Mills 
as well as in supersymmetric
theories.
In four dimensions, by contrast, these notions do not coincide.
To clarify the relationship,
we discuss theories obtained by restricting nonperturbative
sectors.  These theories violate cluster decomposition,
but we illustrate how they may
at least in special cases be understood as disjoint sums of 
well-behaved quantum field theories,
and how dyon spectra can be used to distinguish, for example,
an $SO(3)$ theory with a restriction on instantons from an $SU(2)$ theory.
We also briefly discuss how coupling various analogues of Dijkgraaf-Witten
theory, as part of a description of instanton restriction via coupling
TQFT's to QFT's, may modify these results.

\newpage

\tableofcontents

\newpage

\section{Introduction}

This paper concerns the relationship between gauge theories in which the
gauge group has a center that acts trivially on the matter, and theories with
restrictions on instantons, and applies the relationship to generate
some identities for gauge theories in various dimensions.

In more detail, a $G$ gauge theory where $G$ is semisimple with center $C$
will typically have fewer instantons than a corresponding $G/C$ gauge theory.
A prototype is the relationship between $SU(2)$ and $SO(3)$ gauge theories;
the latter contain instantons that do not lift to $SU(2)$.

Now, although an $SU(2)$ gauge theory contains fewer instantons, it need
not be the same as an $SO(3)$ theory with a restriction on instantons.
The latter, by a well-known result of Weinberg, violates cluster decomposition,
whereas an $SU(2)$ gauge theory by itself will not, at least in four
dimensions.

Thus, we have two related but in general distinct notions:
\begin{itemize}
\item A gauge theory with a trivially-acting subgroup,
\item A gauge theory with a restriction on instantons.
\end{itemize}

In the special case of two dimensions, we will argue that these notions
coincide.  This is a natural generalization of previous work on
orbifolds and abelian gauge theories, encapsulated in the
`decomposition conjecture'
formulated in \cite{summ}.  For orbifolds of the form $[X/\Gamma]$,
for example,
for a finite group $\Gamma$, this says that if a nontrivial subgroup of $\Gamma$
acts trivially on the theory, the CFT of the orbifold is equivalent to 
a disjoint union of CFT's for effective orbifolds.  Projection operators onto
the various components are formed from linear combinations of the
dimension-zero twist fields and the identity operator.

Since there are no gauge dynamics in two dimensions, one would expect
analogous behaviors in nonabelian gauge theories, and that is what we argue
here.  Specifically, for a $G$ gauge theory, $G$ semisimple,
with $K$ a finite subgroup
of the center, if the
matter of the theory is invariant under $K$, then we argue that the theory
decomposes as a disjoint union of $G/K$ gauge theories with variable
discrete theta angles.  For example, schematically,
\begin{displaymath}
SU(2) \: = \: SO(3)_+ \: + \: SO(3)_-.
\end{displaymath}
This can be explicitly checked in a number of
examples.  For one, we describe how Migdal's exact solution of two-dimensional
pure Yang-Mills decomposes as above, as well as how the exact partition
functions of supersymmetric gauge theories on $S^2$ also decompose (when
they have center-invariant matter).

Phrased another way, we will argue in this paper that the decomposition
conjecture of \cite{summ}, which does not specify presentations, does indeed
extend to two-dimensional nonabelian gauge theories, as expected.

In four dimensions, these notions are distinct.
An $SU(2)$ gauge theory is simply not the same as a sum of two $SO(3)$ 
theories, for example.  One can still enforce a restriction on instantons
in an $SO(3)$ theory, but the resulting physical theory differs from
an $SU(2)$ theory.  We will discuss how to see this distinction explicitly
in the spectrum of {\it e.g.} dyons in the theory.

The first half of this paper, section~\ref{sect:2d}, is devoted to
studying examples of two-dimensional gauge theories.  We begin with
a review of the application of the
decomposition conjecture to orbifolds and abelian
gauge theories, which describes how existence of a trivially-acting
finite subgroup of the orbifold or gauge group implies that the 
theory `decomposes' into a disjoint union of mutually non-interacting
ordinary theories.  As gauge fields in two dimensions do not have
propagating degrees of freedom, one would expect the same behavior
in nonabelian gauge theories with trivially-acting finite subgroups,
which we check in examples.  The precise statement of decomposition
for nonabelian gauge theories utilizes two-dimensional
discrete theta angles, so we review those before discussing 
general cases.  We then discuss pure nonsupersymmetric
two-dimensional Yang-Mills as a prototype, and observe how Hilbert spaces,
partition functions, and Wilson line vevs factorize in precisely the
fashion indicated by decomposition.  We also discuss how decomposition
can be seen in theories with center-invariant matter, first via Higgsing
to abelian gauge theories, and then by examining exact expressions
for supersymmetric partition functions on $S^2$.

In section~\ref{sect:4d} we turn to four dimensional theories.
Here, gauge theories with trivially-acting finite subgroups are not
equivalent to gauge theories with restrictions on nonperturbative
sectors.  For example, the latter will violate cluster decomposition
whereas the former need not, unlike two-dimensional cases.
It has been argued that four dimensional SCFT's with restrictions
on instantons may obey an analogue of the decomposition conjecture, 
decomposing into a disjoint union of theories, though much less work
has been done in four dimensions than two.
To better understand both four dimensional decomposition and the distinction
above, we examine dyon spectra
in theories with restrictions on nonperturbative sectors, and observe
how they can be used to distinguish, for example, an $SU(2)$ gauge theory
from an $SO(3)$ gauge theory with a restriction on instantons obeying an
analogue of decomposition.
We also discuss Vafa-Witten topological field theory partition functions
in theories with a restriction on instantons, which seem to provide
a simple example of four dimensional decomposition in action.

Finally, in section~\ref{sect:dw} we discuss Dijkgraaf-Witten theories,
as they are sometimes described in the literature alongside certain
constructions of theories with restrictions on instantons.
Coupling to analogues of 
Dijkgraaf-Witten theory modifies (sometimes, altogether removes)
the decomposition property that much of this paper is devoted to,
as previously observed in two dimensions in \cite{summ}.
We briefly review Dijkgraaf-Witten theory in this context and outline
its effects.

Recently there has been a tremendous amount of progress in understanding
two-dimensional supersymmetric gauge theories and gauged linear sigma models,
one of the central tools used to study string compactifications, and dualities
therein (see for example \cite{ggp1,kl1,jsw,kl2,bpv} for 
a few recent contributions).  
It is our hope that the results presented here on two-dimensional
nonabelian gauge theories will be useful to that effort.

\section{Two dimensional gauge theories}   \label{sect:2d}

Briefly, in two-dimensional theories, we claim that a gauge theory
with a trivially-acting finite subgroup is equivalent to a disjoint union of
theories with effectively-acting gauge groups, unlike 
four dimensional gauge theories.

For orbifolds and
abelian gauge theories in two dimensions, this phenomenon
has by now been extensively documented, and also used to make predictions
for Gromov-Witten invariants, predictions which have since been
checked.
In this context the result has been
known as the `decomposition conjecture' \cite{summ}.

Because in two dimensions gauge fields have no propagating degrees of
freedom and hence no gauge dynamics, surely the same decomposition 
conjecture that applies to orbifolds and abelian gauge theories must
also apply to two-dimensional nonabelian gauge theories.
Briefly, if the center of the gauge group acts trivially on the
massless matter, then the theory contains a trivially-acting orbifold,
and the analysis for orbifolds implies that the theory should
decompose.

We begin by reviewing the application of the
decomposition conjecture to two-dimensional 
orbifolds and abelian gauge
theories, and then turn to nonabelian gauge theories in
two dimensions, where we will see analogous phenomena.
The different factors appearing in the decomposition will have different
discrete theta angles, so we then review of discrete theta
angles, and formulate the general decomposition claim for nonabelian
gauge theories in two dimensions.  We check the decomposition conjecture in
detail for pure nonsupersymmetric Yang-Mills theory in two dimensions, then
discuss checks when center-invariant matter is present, and finally check
via partition functions for (2,2) supersymmetric theories (obtained via
localization).

\subsection{Review of orbifolds and abelian gauge theories}

For two-dimensional orbifolds and abelian gauge theories, it is by now
well-understood that existence of a trivially-acting subgroup is physically
equivalent to a restriction on nonperturbative sectors.  Such theories
break\footnote{
In passing, note that cluster decomposition and locality are different
concepts.  For example, \cite{hs}[appendix B] shows how two-dimensional
BF theory at level $k>1$, which is manifestly local, violates
cluster decomposition.
} cluster decomposition, but in a very mild fashion which is remedied
by the observation that the theory decomposes into a disjoint union of
physical theories with no mutual interactions, indexed by irreducible 
representations of the trivially-acting subgroup.  This has been extensively
discussed and numerous examples described in {\it e.g.} 
\cite{nr,msx,glsm,summ,cdhps,hs,bgmru,addss,hori2,hkm,es-rflat,hetgerbe}.
(See also \cite{me-vienna,me-tex,me-qts} for reviews, 
and \cite{ajt1,ajt2,ajt3,t1,gtseng1,xt1} for checks of 
applications to Gromov-Witten theory).

This behavior is very different from four dimensions:  there,
an $SU(2)$ gauge theory has instantons that are a subset of those of
an $SO(3)$ gauge theory, but the $SU(2)$ gauge theory does not itself
necessarily decompose into a disjoint union of physical theories.
We will explore such relationships in four
dimensions in section~\ref{sect:4d}.

As this behavior in two dimensions has been extensively discussed
elsewhere, in this section we will merely give a brief overview
of the highlights, in order to make this paper somewhat self-contained.

Perhaps the simplest example of this phenomenon is provided by
an orbifold in which one quotients a trivial group action on a space.
As a trivial example, consider $[X/{\mathbb Z}_n]$ where the 
${\mathbb Z}_n$ acts trivially on the entire space.  In this case,
all of the points on $X$ are fixed under the orbifold group action, and
the usual prescription\footnote{
The reader might well be concerned that the usual prescription breaks down
in such a case.  However, it was shown in {\it e.g.} \cite{nr} that
for example modular invariance requires the inclusion of all twisted sectors
of this form in general.
} yields a total of $n-1$ dimension zero twist fields
and a spectrum consisting of $n$ copies of the cohomology of $X$.
As this spectrum contains multiple dimension zero operators, it violates
cluster decomposition, but in the mildest possible way:
this spectrum is equivalent to that of a theory on a disjoint union of
$n$ copies of $X$.  In that description, projection operators 
onto the various components of the disjoint union
are formed from discrete Fourier transforms of the dimension zero operators,
both in this and more general orbifolds of this form.
As a result, correlation functions factorize, both in this and more general
orbifolds of this form.  

A less trivial example is given by the orbifold $[X/D_4]$, where
$D_4$ is the eight-element group with center ${\mathbb Z}_2$.
The elements are given by
\begin{displaymath}
\{1, z, a, b, az, bz, ab, ba=abz\},
\end{displaymath}
where $z$ generates the center of $D_4$.  The quotient $D_4/{\mathbb Z}_2$
is ${\mathbb Z}_2 \times {\mathbb Z}_2$, whose elements we will represent
as
\begin{displaymath}
\{ 1, \overline{a}, \overline{b}, \overline{ab} \},
\end{displaymath}
where, for example, $\overline{a}$ represents the coset $\{a, az\}$.
Now, the one-loop partition function of the $D_4$ orbifold is of the form
\begin{displaymath}
Z(D_4) \: = \: \frac{1}{|D_4|} \sum_{g,h\\gh=hg} Z_{g,h},
\end{displaymath}
where $Z_{g,h}$ denotes the contribution from sectors with boundary
conditions determined by the commuting pair $(g,h)$. 
Now, as the ${\mathbb Z}_2$ acts trivially on $X$, for each pair
$(g,h)$ that appears in the orbifold, their contribution matches that of
a ${\mathbb Z}_2\times {\mathbb Z}_2$ orbifold:
\begin{displaymath}
Z_{g,h} \: = \: Z_{\overline{g},\overline{h}}.
\end{displaymath}
However, not every commuting pair in ${\mathbb Z}_2 \times {\mathbb Z}_2$
lifts to a commuting pair in $D_4$ -- as expected on general principles,
the $D_4$ orbifold has fewer nonperturbative sectors ({\it i.e.}
twisted sectors) than the
${\mathbb Z}_2 \times {\mathbb Z}_2$ orbifold.  Specifically, there are no
$D_4$ twisted sectors of the form
\begin{displaymath}
{\scriptstyle a} \square_b \, , \: \: \:
{\scriptstyle a} \square_{ab} \, , \: \: \:
{\scriptstyle b} \square_{ab} \, ,
\end{displaymath}
as the pairs $(a,b)$, $(a,ab)$, $(b,ab)$ do not commute in $D_4$.
(However, the omitted sectors close under $SL(2,{\mathbb Z})$, so there
is no violation of modular invariance.)
Thus, first, the $D_4$ orbifold cannot match the ${\mathbb Z}_2 \times
{\mathbb Z}_2$ orbifold, as the partition function of the former is
missing some of the sectors of the latter.  Furthermore, the sectors that
are missing, are the same as the ones that acquire signs when one turns
on discrete torsion in a ${\mathbb Z}_2 \times {\mathbb Z}_2$ orbifold.
In fact, with a little work, one can show
\begin{displaymath}
Z(D_4) \: = \: Z({\mathbb Z}_2 \times {\mathbb Z}_2) \: + \:
Z({\mathbb Z}_2 \times {\mathbb Z}_2, \mbox{with discrete torsion}),
\end{displaymath}
the essential point being that the problematic sectors cancel out and so
do not appear in the $D_4$ orbifold.

In open strings, this decomposition can be seen more directly on the
open string states.  A subgroup of the orbifold group that acts
trivially on the target space can act nontrivially on the Chan-Paton
factors.  However, just from taking invariants, the only contributions
that can arise in the open string spectrum are from open strings
connecting Chan-Paton factors in the same representation of the
subgroup in question.  In this way one quickly sees a natural
decomposition of the theory, indexed by irreducible representations of the 
pertinent subgroup, with no open string interactions between different
sectors. 

So far we have discussed orbifolds, but analogous results also hold
in abelian gauge theories in two dimensions with massless states of
nonminimal charge.  One prototype for such considerations is an analogue
of the ordinary (2,2) supersymmetric ${\mathbb C}{\mathbb P}^{N-1}$ model,
with gauge group $U(1)$ and $N$ chiral superfields of charge $k > 1$,
instead of charge $1$ (so that after Higgsing, a ${\mathbb Z}_k$ survives
to act, trivially, on the fields).  
Now, perturbatively, minimal and nonminimal charges
are indistinguishable -- one can rescale one to get the other -- but 
nonperturbatively, the resulting theories can be different \cite{nr}:
\begin{itemize}
\item On a compact worldsheet, to uniquely specify the matter fields,
one must specify a vector bundle to which they couple.  The invariant
meaning of different charge assignments is in terms of different bundles,
and different bundles imply different zero modes, hence different
anomalies, and so different physics.
\item On a noncompact worldsheet, a $U(1)$ gauge theory with minimal
charges can be distinguished from one with nonminimal charges by the existence
of massive minimally-charged fields.  Even if their masses are above
the cutoff scale in the theory, their presence can still be detected
via periodicity of the two-dimensional $\theta$ angle.
\end{itemize}

Such abelian gauge theory examples containing a trivially-acting
finite subgroup also obey decomposition.
As a concrete example, consider a generalization of the
${\mathbb C}{\mathbb P}^{N-1}$ example above, a GLSM with $N$ chiral superfields
$x_i$, another chiral superfield $z$, and gauge group $U(1)^2$ with
charges
\begin{center}
\begin{tabular}{cc} 
 $x_i$ & $z$ \\ \hline
$1$ & $-n$ \\
$0$ & $k$
\end{tabular}
\end{center}
where $n$ is positive.  This theory has D-terms
\begin{eqnarray*}
\sum_i |x_i|^2 \: - \: n |z|^2 & = & r_1, \\
k |z|^2 & = & r_2,
\end{eqnarray*}
where we assume $r_1 \gg 0$, $r_2 \gg 0$.
This example is discussed in \cite{glsm}[section 3.3].
The second $U(1)$ nearly gauges away the $z$, except for a remaining
trivial ${\mathbb Z}_k$ action.  The remaining fields and $U(1)$
describe the ${\mathbb C} {\mathbb P}^{N-1}$ model.  The result\footnote{
Technically, this is the GLSM for a ${\mathbb Z}_k$ gerbe on
${\mathbb C}{\mathbb P}^{N-1}$, of characteristic class $-n$ mod $k \in
H^2({\mathbb C}{\mathbb P}^{N-1},{\mathbb Z}_k)$.
} is
${\mathbb C}{\mathbb P}^{N-1}$ with a trivial ${\mathbb Z}_k$
action, classified by $n$ mod $k$.
The quantum cohomology ring of this theory was shown in 
\cite{glsm}[section 3.3] to be
\begin{displaymath}
{\mathbb C}[x,y] / (x^N y^{-n} = q, y^k = 1).
\end{displaymath}
The different values of $y$ index $k$ different copies of 
the quantum cohomology ring of ${\mathbb C}{\mathbb P}^{N-1}$,
with shifting $B$ fields.  Thus, the quantum cohomology ring of
this theory is consistent with decomposition.

The considerations above also apply to nonlinear sigma models in two 
dimensions.  If we restrict nonperturbative sectors, the resulting
theory violates cluster decomposition, but in an extremely mild way,
as the result is equivalent to a disjoint union of ordinary theories
with variable $B$ fields.
Schematically, we can see this in the path integral for a nonlinear
sigma model as follows.  To project onto sectors with worldsheet instantons
defined by maps $\phi$ such that 
\begin{displaymath}
\int \phi^* \omega
\end{displaymath}
is divisible by $n$, for some fixed two-form $\omega$, for example,
is accomplished by inserting a projection operator of the form
\begin{displaymath}
\int [D \phi] e^{-S} \left( \sum_{k=0}^{n-1} e^{2 \pi i (k/n)\int \phi^* \omega}
\right),
\end{displaymath}
which is the same as
\begin{displaymath}
\sum_{k=0}^{n-1} \int [D \phi] \exp\left( -S \: + \: 2 \pi i (k/n)
\int \phi^* \omega \right).
\end{displaymath}
In this second form, the interpretation as a disjoint union of theories
with rotating $B$ fields (defined by $(k/n) \omega$) is clear.

The decomposition conjecture \cite{summ} gives
a very precise description of how to build the disjoint union from
the original nonlinear sigma model with a restriction on instantons.
Such theories also give physical realizations of strings on certain stacks
known as gerbes, and so the decomposition conjecture
makes physical predictions for Gromov-Witten
invariants of strings on gerbes.  These predictions have been
rigorously proven,
see for example \cite{ajt1,ajt2,ajt3,t1,gtseng1,xt1}.

\subsection{Discrete theta angles}

Just as in four dimensions \cite{gmn,ast}, there is also a discrete theta angle
in two dimensions, as has been observed in 
\cite{hori94a,hori94b,hori2,yuji13}.
These discrete theta angles play an essential role in the form of
decomposition for two-dimensional nonabelian gauge theories, so we review
them here.

Consider a $G$ gauge theory with $G = \tilde{G}/K$, 
$\tilde{G}$ semisimple and simply-connected, 
and 
$K$ a finite subgroup of the center of $\tilde{G}$.
$K$ will be a product of cyclic groups, {\it i.e.}
\begin{displaymath}
K \: = \: {\mathbb Z}_{n_1} \oplus {\mathbb Z}_{n_2} \oplus \cdots .
\end{displaymath}
There is a degree two
$K$-valued characteristic class $w$, which for $SO(n)$ for
example would be the degree two Stiefel-Whitney class.
Each such characteristic class can be used to define a discrete theta
angle via a term in the action which is schematically\footnote{
To be clear, such finite-group-valued cohomology does not have a representation
in de Rham cohomology, so we are using an integral loosely to denote a
formal contraction.
}
of the form
\begin{displaymath}
\theta' \int w 
\end{displaymath}
for $\theta'$ a character\footnote{
As $w$ is $K$-valued, one contracts with a character of $K$ to derive
a number.
} of $K$.

Thus, for example, in two dimensions there are two $SO(3)$ gauge
theories, which we shall label $SO(3)_{\pm}$, just as in four dimensions.

In general, the set of discrete theta angles for a gauge group $G$
as above is indexed by the set $\hat{K}$ of irreducible representations of
$K$ (which, since $K$ is a finite abelian group, has as many elements as
$K$, and since $K$ is a sum of cyclic groups, can be identified with the
characters of $K$).  
Thus, we will index discrete theta angles in $G$ gauge theories
by\footnote{
As a minor consistency check, note that
the components of the decomposition in
orbifolds and abelian gauge theories described in \cite{summ} were also
indexed by irreducible representations.  Here, this is a consequence of the
construction of discrete theta angles, whereas in \cite{summ} the 
justification was completely different, relying instead on {\it e.g.}
observations on open string spectra.
} $\mu \in \hat{K}$.

Furthermore, if $G$ is any semisimple group and $K$ a finite subgroup of its 
center, then a $G/K$ gauge theory will contain $|K| = |\hat{K}|$ more
discrete theta angles than $G$.  These new discrete theta angles will not
be entirely independent of the old ones.  For example, 
$(SU(4)/{\mathbb Z}_2)/{\mathbb Z}_2 = SU(4)/{\mathbb Z}_4$ has two 
more discrete theta angles relative to $SU(4)/{\mathbb Z}_2$, which itself
has two, so that $SU(4)/{\mathbb Z}_4$ has four altogether.
More generally, letting $C$ denote
centers, if $G = \tilde{G}/K$ then there is an extension
\begin{equation}   \label{eq:gp-ext-1}
1 \: \longrightarrow \: K \: \longrightarrow \: C(\tilde{G}) \: \longrightarrow
\: C(G) \: \longrightarrow \: 1,
\end{equation}
and hence the discrete theta angles are related by\footnote{
A few details may be useful for some readers.  The dual sequence above is
obtained by applying Hom$(-,U(1))$, and so one gets a sequence
\begin{displaymath}
1 \: \longrightarrow \: \widehat{ C(G) } \: \longrightarrow \:
\widehat{ C(\tilde{G}) } \: \longrightarrow \: \hat{K} \:
\longrightarrow \: {\rm Ext}^1(C(G),U(1)) .
\end{displaymath}
In the cases at hand, $C(G)$ will always be a sum of cyclic groups, and
Ext$^1({\mathbb Z}_n,U(1)) = 0$, hence Ext$^1(C(G),U(1)) = 0$.  To see this,
we apply Hom$(-,U(1))$ to the sequence
\begin{displaymath}
0 \: \longrightarrow \: {\mathbb Z} \: \stackrel{\times n}{\longrightarrow} \:
{\mathbb Z} \: \longrightarrow \: {\mathbb Z}_n \: \longrightarrow \: 0
\end{displaymath}
to get
\begin{displaymath}
{\rm Hom}({\mathbb Z},U(1)) \: \longrightarrow \: {\rm Hom}({\mathbb Z},U(1))
\: \longrightarrow \: {\rm Ext}^1({\mathbb Z}_n,U(1)) \: \longrightarrow \: 0
\end{displaymath}
and use Hom$({\mathbb Z},U(1)) = U(1)$, hence the first map is the surjective
$n$th power map
and Ext$^1({\mathbb Z}_n,U(1)) = 0$.  We would like to thank S.~Katz for
the computation shown here.
}
\begin{equation}  \label{eq:char-ext-1}
1 \: \longrightarrow \: \widehat{ C(G) } \: \longrightarrow \:
\widehat{ C(\tilde{G}) } \: \longrightarrow \: \hat{K} \:
\longrightarrow \: 1 .
\end{equation}

\subsection{Decomposition conjecture for nonabelian gauge theories}
\label{sect:decomp-conj}

Briefly, if a nonabelian gauge theory with semisimple gauge group $G$ has a 
center $C(G)$ that acts trivially on the massless matter of the theory, then 
this theory implicitly contains a trivially-acting $C(G)$ orbifold, and so
ought to admit the same decomposition structure described earlier.
We shall check this assertion in detail, utilizing exact solutions for
pure nonsupersymmetric Yang-Mills theory and recent localization-derived
exact results for supersymmetric gauge theories in two dimensions.

To set notation, let $\tilde{G}$ denote a semisimple,
simply-connected, compact Lie group.
We will compare $\tilde{G}$ gauge theories with center-invariant matter
to $\tilde{G}/C(\tilde{G})$ gauge theories with the same matter,
where $C(\tilde{G})$ denotes the center.  The latter theories have
(finitely many) discrete theta angles, indexed by irreducible representations
of $C(\tilde{G})$.  Letting a theory with discrete theta angle
indexed by $\mu \in \widehat{ C(\tilde{G}) }$ be denoted
\begin{displaymath}
\left( \tilde{G}/C(\tilde{G}) \right)_{\mu}
\end{displaymath}
we will argue that the gauge theories decompose as
\begin{equation}   \label{eq:decomp-simple}
\tilde{G} \: = \: \sum_{\mu \in \widehat{C(\tilde{G})}} \left(
\tilde{G}/C(\tilde{G}) \right)_{\mu} .
\end{equation}
For example,
\begin{displaymath}
SU(2) \: = \: SO(3)_{+} \: + \: SO(3)_-
\end{displaymath}
and
\begin{displaymath}
SU(4) \: = \: (SU(4)/{\mathbb Z}_4)_0 \: + \: (SU(4)/{\mathbb Z}_4)_1 
\: + \: (SU(4)/{\mathbb Z}_4)_2 \: + \: (SU(4)/{\mathbb Z}_4)_3 .
\end{displaymath}

A little more generally, for any semisimple compact $G$, not necessarily
simply-connected, and for $K$ a subgroup of the center, we have
discrete theta angles in $\hat{K}$ and a decomposition
\begin{equation}  \label{eq:decomp-simple2}
G \: = \: \sum_{\lambda \in \hat{K}} (G/K)_{\lambda} .
\end{equation}
Now, in this case, $G$ itself may also have discrete theta angles, so
we can generalize to include the decomposition of theories with
discrete theta angles.

Generalizing along such lines,
let $G = \tilde{G}/K$ for $\tilde{G}$ as above and
$K$ some subgroup of $C(\tilde{G})$,
where $G$ has center $C(G)$.
Then, $G$ gauge theories are indexed by a discrete theta angle
$\lambda \in \hat{K}$.  In this case, we can decompose $G$ gauge theories
with discrete theta angle $\lambda$ as
\begin{equation}    \label{eq:decomp-general}
( G )_{\lambda} \: = \: 
\sum_{\mu \in \widehat{C(G)} } \left( G/C(G) \right)_{\lambda,\mu} .
\end{equation}

Since 
\begin{displaymath}
G/C(G) \: = \: \tilde{G} / C(\tilde{G})
\end{displaymath}
we can interpret $(\lambda \in \hat{K}, \mu \in \widehat{C(G)})$
as defining an element\footnote{
There are noncanonical isomorphisms of sets $\hat{K} \times \widehat{ C(G) }
\rightarrow \widehat{ C(\tilde{G}) }$.  Unless $\widehat{ C(\tilde{G}) }$
splits, however, the product structure will be more complicated. 
} of $\widehat{ C(\tilde{G})}$.  To be somewhat more specific,
we will choose always an isomorphism that respects the 
extension~(\ref{eq:gp-ext-1}), by which we mean, if 
$g \in C(\tilde{G})$ is in the image of $K$, then we define $(\lambda,\mu)$
to be such that
\begin{equation}   \label{eq:prod-defn}
(\lambda, \mu)(g) \: = \: \lambda(g).
\end{equation}

As a consequence, these decompositions are related as follows:
\begin{equation}   \label{eq:decomp-summary}
\tilde{G} \: = \: \sum_{\lambda \in \hat{K}} (G)_{\lambda} \: = \:
\sum_{\rho \in \widehat{ C(\tilde{G}) } } ( \tilde{G}/C(\tilde{G}) )_{\rho} .
\end{equation}

We list a few examples of the predictions of this more
general form of the decomposition conjecture below.
First, using the fact that
\begin{displaymath}
SO(4) \: = \: \frac{ SU(2) \times SU(2) }{ {\mathbb Z}_2 }
\end{displaymath}
we have
\begin{eqnarray*}
SO(4)_+ & = & 
SO(3)_+ \times SO(3)_+ \: + \: SO(3)_- \times SO(3)_- ,\\
SO(4)_- & = & 
SO(3)_+ \times SO(3)_- \: + \: SO(3)_- \times SO(3)_+ .
\end{eqnarray*}

Note as a consequence that
\begin{eqnarray*}
SU(2) \times SU(2) & = &
SO(4)_+ \: + \: SO(4)_- ,\\
& = & \sum_{i,j=\pm} SO(3)_i \times SO(3)_j ,
\end{eqnarray*}
as expected from equation~(\ref{eq:decomp-summary}).

Another example follows from the fact that the center of $SU(4)$ is
${\mathbb Z}_4$.  If instead we take a ${\mathbb Z}_2$ quotient, then
we get
\begin{eqnarray*}
\left( SU(4)/{\mathbb Z}_2 \right)_+ & = &
\left( SU(4)/{\mathbb Z}_4 \right)_0 \: + \:
\left( SU(4)/{\mathbb Z}_4 \right)_2 ,
\\
\left( SU(4)/{\mathbb Z}_2 \right)_- & = &
\left( SU(4)/{\mathbb Z}_4 \right)_1 \: + \:
\left( SU(4)/{\mathbb Z}_4 \right)_3 .
\end{eqnarray*}
The result above follows from the fact that ${\mathbb Z}_4 / {\mathbb Z}_2$
contains two cosets, $\{0,2\}$ and $\{1,3\}$.
Note that, as a consequence,
\begin{eqnarray*}
SU(4) & = & \left( SU(4)/{\mathbb Z}_2 \right)_+ \: + \:
\left( SU(4)/{\mathbb Z}_2 \right)_- ,\\
& = & \sum_{k=0}^3 \left( SU(4)/{\mathbb Z}_4 \right)_k ,
\end{eqnarray*}
as expected from equation~(\ref{eq:decomp-summary}).

\subsection{Decomposition in nonsupersymmetric pure gauge theories}

In this section we will verify the decomposition claim for 
nonsupersymmetric pure Yang-Mills theories.  For theories with
vanishing discrete theta angles, exact expressions for Hilbert spaces,
partition functions, and so forth are known (see for example
\cite{migdal1,rusakov1,gt1,cmr,witten-2dgauge}).  To verify the
decomposition claim, we will utilize extensions of those
results to theories with discrete theta angles.  Such extensions
have been discussed previously in \cite{yuji13}, and we will 
discuss and elaborate on them here\footnote{
We would like to thank Y.~Tachikawa for discussions of his results on
two-dimensional nonsupersymmetric pure Yang-Mills theories with
discrete theta angles.
}.

Let us begin with a discussion of the Hilbert spaces of 
nonsupersymmetric pure Yang-Mills theory with gauge group
$G = \tilde{G}/K$, where $\tilde{G}$ is compact, semisimple,
and simply-connected, and $K$ is
a (finite) subgroup of the center of $\tilde{G}$.

For the theory with vanishing discrete theta angles, the conventional
gauge theory, the Hilbert space is the space of functions $f(g)$ on
$G$, invariant under conjugation.  Such functions, known as class
functions, can be
expanded in an analogue of a Fourier series in characters $\chi$ of $G$, as
\begin{displaymath}
f(g) \: = \: \sum_R c_R \chi_R(g)
\end{displaymath}
for constants $c_R$ determined by the function $f$.

Now, let us consider the Hilbert space of the corresponding theory with
a nonzero discrete theta angle defined by $\lambda \in \hat{K}$.
Because of the discrete theta angle term in the Lagrangian, a particle
moving around a closed noncontractible path must pick up a phase.
Thus, we should think about the Hilbert space as consisting of sections of
a line bundle on $G$,
or equivalently a class function $f$ on $\tilde{G}$ satisfying
\begin{displaymath}
f(g z) \: = \: \lambda(z) f(g ) ,
\end{displaymath}
where $z \in K$.
As a result, $f(g)$ can be expanded in terms of characters $\chi_R$
of $\tilde{G}$ which
are in a fixed representation of $K$.

Before going on to describe partition functions, let us briefly describe
how the result above is compatible with decomposition.  Recall from
equation~(\ref{eq:decomp-simple2}) that
decomposition predicts that for pure nonsupersymmetric Yang-Mills theories
with semisimple gauge group $G$ and $K$ a (finite) subgroup of the center,
the theory should decompose in a fashion we indicate schematically as
\begin{displaymath}
G \: = \: \sum_{\lambda \in \hat{K}} (G/K)_{\lambda} .
\end{displaymath}
Now, the Hilbert space of the $G$ gauge theory contains all
class functions on $G$.  On the left-hand-side, the Hilbert space
of each $( G/K )_{\lambda}$ theory contains class functions of $G$ which
are in a fixed representation of $K$.  The set of all class functions on
$G$ has a natural decomposition according to representations of $K$, and so
we see that the Hilbert spaces match in exactly the fashion predicted
by decomposition.

In addition, as $G$ gauge theories may themselves have discrete theta
angles, decomposition makes a more refined conjecture~(\ref{eq:decomp-general}),
\begin{displaymath}
( G )_{\lambda} \: = \: 
\sum_{\mu \in \widehat{C(G)} } \left( G/C(G) \right)_{\lambda,\mu}
\end{displaymath}
for $G = \tilde{G}/K$, $K$ a subgroup of $C(\tilde{G})$.
In terms of the Hilbert space under discussion, this reflects the fact
since $K$ commutes with $C(\tilde{G})$, class functions on
$\tilde{G}$ are simultaneously representations of $K$ and $C(\tilde{G})$.
In particular, the Hilbert space on the
left, consisting of class functions in representation $\lambda$ of $K$,
can be further decomposed according to representations of $C(\tilde{G})$.

A brief example may help illuminate this matter.  Suppose
$\tilde{G} = SU(4)$ and $K = {\mathbb Z}_2$, so $G = SU(4)/{\mathbb Z}_2$.
The Hilbert space of a pure $G$ gauge theory of the form discussed
here then decomposes into class functions of fixed
${\mathbb Z}_2$ representations.  However, each ${\mathbb Z}_2$
representation can be further decomposed into ${\mathbb Z}_4$
representations.  If we let $n \in \{0, 1, 2, 3\}$ characterize
representations of ${\mathbb Z}_4$, then, schematically,
\begin{center}
\begin{tabular}{ccc}
trivial rep' of ${\mathbb Z}_2$ & $\leftrightarrow$ &
0, 2 of ${\mathbb Z}_4$ ,\\
nontrivial rep' of ${\mathbb Z}_2$ & $\leftrightarrow$ &
1, 3 of ${\mathbb Z}_4$ .
\end{tabular}
\end{center}
In this fashion, we see that Hilbert spaces reproduce the
decomposition~(\ref{eq:decomp-general}).

Now, let us turn to partition functions.
In two-dimensional pure gauge theories with vanishing discrete theta
angles, the partition functions are
known exactly \cite{migdal1,rusakov1,gt1}, and are of the form
(\cite{cmr}[equ'n (3.20)], \cite{witten-2dgauge}[equ'n (2.51)])
\begin{displaymath}
Z \: = \: \sum_R ({\rm dim}\, R)^{2-2g} \exp\left( - A
C_2(R) \right) ,
\end{displaymath}
where $g$ is the genus of the two-dimensional spacetime, $A$ its
area, and the sum is over representations of the gauge group.

Now, let us work out partition functions of two-dimensional theories with
discrete theta angles. To that end, it is helpful to consider a 
genus-one surface with one end sliced off, as illustrated below:
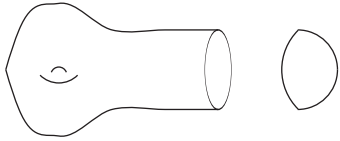
\begin{center}
\begin{picture}(155,100)(0,0)
\CArc(120,50)(15,270,360) \CArc(120,50)(15,0,90)
\CArc(135,50)(21,135,225)
\Oval(90,50)(15,5)(0)
\CArc(30,55)(10,225,315)
\CArc(30,48)(3,20,160)
\Curve{(90,65)(70,65)(50,65)(30,75)(15,65)(10,50)}
\Curve{(10,50)(15,35)(30,25)(50,35)(70,35)(90,35)}
\end{picture}
\end{center}
Since the wavefunctions around noncontractible loops are, as discussed
above, understood as sections of nontrivial bundles over $G$, the partition
function of the left slice must also be such a section.  Since the entire
partition function is obtained by gluing, and the contribution of the
left-hand-side is a section of a nontrivial bundle, the contribution from
the cap on the right must also be a section of the same nontrivial bundle. 

Such sections of nontrivial bundles can be understood in terms of
class functions on $\tilde{G}$ associated to specific $\mu \in \hat{K}$,
as previously discussed.  The resulting partition function, for
a theory with discrete theta angle defined by $\mu \in \hat{K}$, should be
of the same form as before, namely
\begin{displaymath}
Z \: = \: \sum_R ({\rm dim}\, R)^{2-2g} \exp\left( - A
C_2(R) \right) ,
\end{displaymath}
except that here, the sum over representations is restricted to
representations $R$ of $G$ associated with $\mu$.
(This result has previously been given in \cite{yuji13}.)

The result above is invariant under retriangulations, by virtue of nearly
identical computations to cases with vanishing discrete theta angles.
The essential point is that to prove independence from choice of triangulation
and related results
requires the following four identities \cite{cmr,witten-2dgauge}:   
\begin{equation}    \label{eq:2dgauge-orthog}
\int dU \overline{ \chi_{R}(U) } \chi_{R'}(U) \: = \: 
\delta_{R, R'} ,
\end{equation}
\begin{equation}   \label{eq:2dgauge-complete}
\sum_R | \chi_R \rangle \langle \chi_R | \: = \: 1 ,
\end{equation}
\begin{equation}    \label{eq:2dgauge-glue1}
\int dU \chi_R(A U B U^{-1}) \: = \: \frac{1}{{\rm dim}\,R} 
\chi_R(A) \chi_R(B) ,
\end{equation}
\begin{equation}    \label{eq:2dgauge-glue2}
\int d V \chi_R(AV) \chi_{R'}(V^{-1}B) \: = \: \delta_{R,R'} 
\frac{1}{{\rm dim}\, R} \chi_R(AB) .
\end{equation}
Three of the equations above apply automatically without modification.
Only equation~(\ref{eq:2dgauge-complete}) requires any thought.
Because $K$ is a subgroup of the center, $K$ commutes with the gauge
group, and so the sum over representations in 
equation~(\ref{eq:2dgauge-complete}) can be further diagonalized to
provide a set of completeness relations, one for each representation of
$K$.  Thus, after suitable renormalizations, one can write
\begin{displaymath}
\sum_{R, {\rm fixed}\, \mu} | \chi_R \rangle \langle \chi_R | \: = \: 1
\end{displaymath}
for each $\mu \in \hat{K}$.

We can derive results for partition functions axiomatically, as in
\cite{cmr}[section 3.7], from a nearly identical ansatz:
\begin{eqnarray}
\mbox{Cap} & = & \sum_R ({\rm dim}\, R) \chi_R(U) e^{-A C_2(R)} \: \equiv
\: Z_{\rm cap}(U) ,\\
\mbox{Tube} & = & \sum_R | R \rangle \langle R | e^{-A C_2(R) },\\
\mbox{Pants} & = & \sum_R |R\rangle \otimes | R \rangle \otimes
| R \rangle \frac{ e^{-A C_2(R)} }{{\rm dim}\,R} ,
\end{eqnarray}
where in each case, the sum is over representations associated
to fixed $\mu \in \hat{K}$.

In this language, we can understand invariance under retriangulations,
for example, in the following standard fashion.
Consider a cap which has been subdivided down the middle,
as illustrated below:
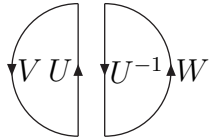
\begin{center}
\begin{picture}(80,70)
\ArrowArc(35,35)(25,90,270)
\ArrowLine(35,10)(35,60)
\ArrowArc(45,35)(25,-90,90)
\ArrowLine(45,60)(45,10)
\Text(33,35)[r]{$U$}  \Text(12,35)[l]{$V$}
\Text(47,35)[l]{$U^{-1}$}  \Text(72,35)[l]{$W$}
\end{picture}
\end{center}
To be independent of triangulation means that if we glue the caps above
along $U$, the result should be a cap in which $U$ does not appear,
{\it i.e.}
\begin{displaymath}
\int dU Z_{\rm cap}(VU) Z_{\rm cap}(U^{-1} W) \: = \:
Z_{\rm cap}(VW)
\end{displaymath}
and this is a nearly immediate consequence of
equation~(\ref{eq:2dgauge-glue2}).  This analysis is nearly identical
to that of \cite{cmr}[section 3.4.2], the only difference being that here
sums over representations are restricted to those representations with fixed
$\mu \in \hat{K}$.  Other arguments from \cite{cmr}[chapter 3] apply
here to two-dimensional Yang-Mills with discrete theta angles with equal
immediacy.

To make the discussion above more clear, let us outline the results for
the $SU(2)$ and $SO(3)_{\pm}$ gauge theories.  For $SU(2)$,
the partition function is given by
\begin{displaymath}
Z \: = \: \sum_R ({\rm dim}\, R)^{2-2g} \exp\left( - A
C_2(R) \right) ,
\end{displaymath}
where the sum is over all representations $R$ of $SU(2)$, and for
$SO(3)_+$, the sum is over all representations of $SO(3)$.
For $SO(3)_-$, the sum is over $SU(2)$ representations that are not
also $SO(3)$ representations.  That sounds somewhat odd as a description
of an $SO(3)$ gauge theory, but the point is that it should be interpreted
in terms of sections of nontrivial bundles over $SO(3)$.

In this case, decomposition should now be clear:  the sum appearing
in the $SU(2)$ gauge theory is the sum of the representations appearing
in the $SO(3)_+$ and $SO(3)_-$ gauge theories, so
the partition functions obey decomposition:
\begin{displaymath}
Z(SU(2)) \: = \: Z(SO(3)_+) \: + \: Z(SO(3)_-) .
\end{displaymath}

More generally, in a $G$ gauge theory, summing over the
various discrete theta angles in $G/K$ gauge theories reproduces the 
sum over $G$ representations, and so the partition functions obey
decomposition~(\ref{eq:decomp-simple2}):
\begin{displaymath}
Z(G) \: = \: \sum_{\lambda \in \hat{K}} Z( (G/K)_{\lambda} ) .
\end{displaymath}

It is not difficult to see how the further 
generalization~(\ref{eq:decomp-general}) also arises.
Recall this form of the decomposition conjecture says, schematically,
\begin{displaymath}
( G )_{\lambda} \: = \: 
\sum_{\mu \in \widehat{C(G)} } \left( G/C(G) \right)_{\lambda,\mu}
\end{displaymath}
for $G = \tilde{G}/K$.  Here, the point is that the partition function
\begin{displaymath}
Z( (G)_{\lambda})
\end{displaymath}
involves a sum over representations of $\tilde{G}$ that are in
a fixed representation of $K$.  As noted in the discussion of Hilbert
spaces, representations of $\tilde{G}$ in a fixed representation of $K$
can be further decomposed into representations of $C(G)$, 
hence the partition functions decompose in the fashion outlined above.

In addition to exact expressions for Hilbert spaces and partition functions,
there also exist exact expressions for vevs of Wilson lines in 
two-dimensional Yang-Mills, see for example  
\cite{cmr}[section 3.5], \cite{bt-ym}[equ'n (3)].  
These expressions also generalize to nonzero
discrete theta angles, and obey a decomposition principle. 

Briefly, closely following the discussion in \cite{cmr}[section 3.5.1] for
nonintersecting loops, if one has a Wilson loop defined by a curve
$\Gamma$ and representation $R_{\Gamma}$,
\begin{displaymath}
W(R_{\Gamma}, \Gamma) \: = \: {\rm Tr}_{R_{\Gamma}} P \exp\left(
\oint_{\Gamma} A \right) ,
\end{displaymath}
then
\begin{displaymath}
\langle W(R_{\Gamma},\Gamma) \rangle \: = \:
\int dU \prod_c Z(\Sigma_c, U, U^{-1} ) W(R_{\Gamma}, \Gamma) ,
\end{displaymath}
where the $\Sigma_c$ are the various components of the two-dimensional
spacetime obtained after slicing along $\Gamma$, with boundaries labelled
by group elements $U$, $U^{-1}$ according to orientation.
As in \cite{cmr}[section 3.5.1], the expression above can be
written as a sum over representations, in essentially the usual form.

From the discussion above, we see that vevs of Wilson lines factorize
in exactly the same fashion as Hilbert spaces and partition functions,
and for the same fundamental reason.  The vev of a Wilson line in a
$(G)_{\lambda}$ theory, for example, for $G = \tilde{G}/K$ and 
$\lambda \in \hat{K}$, is defined by a sum over $\tilde{G}$ representations
associated to the fixed representation $\lambda$ of $K$.  As before, those
representations can be further decomposed into representations of
$C(G)$, from which the usual decomposition statement follows.  As the
details are more or less identical to what has been described previously,
we will not extensively elaborate.  In passing, note that the vev of 
a given Wilson line in some components may vanish -- the sum will add up
to the vev of the Wilson line, but the contributions to the sum need not
be separately nonzero.

Briefly, as one further consistency check, ($q$-deformed)
pure nonsupersymmetric
two-dimensional Yang-Mills was related to $G/G$ gauged WZW models in
\cite{klimcik}.  We will not discuss gauged WZW models in great detail,
but we do note in passing that they necessarily obey their own
analogue of the decomposition conjecture.  Specifically, if $G$ is
a semisimple Lie group and $H \subset G$ a subgroup with center $C$,
then in a $G/H$ gauged WZW model, since one gauges the adjoint action
of $H$, the center $C$ necessarily acts trivially on the theory, so
applying the same reasoning as for orbifolds and abelian gauge theories,
the $G/H$ gauged WZW model must also obey decomposition.   

As another consistency check, pure nonsupersymmetric two-dimensional
Yang-Mills is also closely related to $BF$ theory in two dimensions,
see for example \cite{bt-2dgauge}.  It is therefore relevant to mention
that $BF$ theory in two dimensions at level $k>1$ also exhibits
a breakdown in cluster decomposition, which can be understood in terms
of a decomposition of the theory into disjoint sectors \cite{hs}[appendix B].

Finally, in passing, two-dimensional pure nonsupersymmetric Yang-Mills also
arises in other contexts.  One such is described in
\cite{witten-2dgauge}[section 3.1], in terms of volumes of moduli spaces
of flat connections.  We will not elaborate on such matters here,
but we do note that the decomposition of volumes of moduli spaces of
flat $SO(3)$ connections described there is consistent with the
decomposition conjecture presented here.

\subsection{Decomposition in theories with center-invariant matter}

Let us now consider adding center-invariant matter, to a 
not-necessarily-supersymmetric pure gauge theory in two dimensions.
We claim the decomposition conjecture continues to hold.
We are not aware of exact partition function results for 
non-supersymmetric gauge theories with matter\footnote{
That said, there has been recent progress in solving two-dimensional
Yang-Mills with adjoint-valued matter \cite{katz-qcd}.
}, but we can
perform other checks.

Specifically, we can Higgs the nonabelian gauge symmetry to 
a subgroup, and apply the decomposition conjecture in its form for
orbifolds and abelian gauge theories to recover a decomposition of the
desired form.

First, consider a two-dimensional $SU(2)$ theory containing an adjoint
scalar.  At a generic point on the Higgs branch, the $SU(2)$ has been
Higgsed to $U(1)$.  If the other matter is invariant under the
${\mathbb Z}_2$ center, then its $U(1)$ charges must be even,
as $\{\pm 1\} \subset U(1)$ must leave the matter invariant.

Now, we know from our review of abelian gauge theories that a $U(1)$
gauge theory with massless matter of even charges\footnote{
In purely abelian gauge theories, one also needed to give massive
minimally-charged matter, and then one could use $\theta$ angle
periodicities to distinguish these theories from theories obtained merely
by a rescaling.  Here, equivalent information seems to be implicitly
encoded in the fact that we are Higgsing an $SU(2)$ theory.
} will decompose into
a disjoint union of two theories, with variable flat $B$ fields.
We can understand each of those components as the result of Higgsing
the two $SO(3)$ gauge theories.  In this language, the flat $B$ fields
are the low-energy description of the discrete theta angles.

For another example, consider an $SU(3)$ gauge theory with center-invariant
matter,  As above, consider a point on the Higgs branch where the
$SU(3)$ is Higgsed to a $U(1)^2$ (so that we can apply our understanding
of abelian gauge theories).  Since the center of $SU(3)$ is generated by
\begin{displaymath}
{\rm diag}\left( \xi, \xi, \xi \right)
\end{displaymath}
for $\xi$ a primitive third root of unity, and the $U(1)^2 \subset SU(3)$
is given by
\begin{displaymath}
{\rm diag}\left( e^{i \theta_1}, e^{i \theta_2}, e^{-i(\theta_1+\theta_2)}
\right) ,
\end{displaymath}
we see that if the matter is invariant under the center of $SU(3)$,
then it must be invariant under the subgroup $(1, \xi, \xi^2) \subset
U(1)$ for each $U(1)$, and hence all matter fields must have, after Higgsing,
charges that are multiples of three.  At this point, as before, we can
apply our understanding of decomposition in abelian gauge theories to
argue that, at least along the Higgs branch, the theory decomposes
into a disjoint union of three theories, each of which should be the
result of Higgsing an $SU(3)/{\mathbb Z}_3$ gauge theory with
suitable discrete theta angle.  In this fashion we get another
consistency check of the decomposition conjecture for nonabelian
gauge theories.

More generally, given a $G = \tilde{G}/K$ gauge theory, $\tilde{G}$
semisimple, simply-connected, and compact,
$K$ a subgroup of the center, with matter that
is invariant under $K$, essentially the same considerations apply
on the Higgs branch.  Briefly, if we Higgs\footnote{
If the matter is such that such a Higgsing is not possible, then it is
not possible to reduce to considerations of abelian gauge theories and
apply older results.
} the
gauge group $G$ to a product
of $U(1)$'s, and, because the matter is center-invariant, the $U(1)$
charges must all be nonminimal.  The known version of the decomposition
conjecture then applies to give results consistent with the nonabelian
version described in section~\ref{sect:decomp-conj}.

\subsection{Decomposition of supersymmetric
theories}

For (2,2) supersymmetric gauge theories on $S^2$, there are now
exact results for partition functions \cite{benini1,doroud1}, 
obtained via localization.  In this section, we will show that
those exact partition functions for supersymmetric theories with
center-invariant matter obey the decomposition conjecture stated
previously in section~\ref{sect:decomp-conj}.
We will follow the notation of \cite{benini1}.

Let us begin with a comparison of the supersymmetric partition functions
for the pure $SU(2)$ and $SO(3)_{\pm}$ theories.
Following \cite{benini1}[equ'n (3.34)], the partition function of
a supersymmetric theory on $S^2$ is given by
\begin{equation}   \label{eq:part-fn}
Z_{S^2} \: = \: \frac{1}{|{\cal W}|} \sum_{\mathfrak m} 
\int \left( \prod_j \frac{d \sigma_j}{2\pi} \right) Z_{\rm class}(
\sigma, {\mathfrak m}) Z_{\rm gauge}(\sigma, {\mathfrak m})
\prod_{\Phi} Z_{\Phi}(\sigma, {\mathfrak m}; \tau, {\mathfrak n}) ,
\end{equation}
where \cite{benini1}[equ'n (3.35)]
\begin{eqnarray*}
Z_{\rm class}(\sigma,{\mathfrak m}) 
& = & e^{-4 \pi i \xi {\rm Tr}\, \sigma - i \theta
{\rm Tr} \, {\mathfrak m}} \exp\left( 8 \pi i r
{\rm Re}\, \tilde{W}(\sigma/r + i {\mathfrak m}/(2r)) \right), \\
Z_{\rm gauge}(\sigma,{\mathfrak m}) & = & 
\prod_{\alpha \in G} \left( \frac{ |\alpha( 
{\mathfrak m})| }{2} \: + \: i \alpha(\sigma) \right)
\: = \:
\prod_{\alpha > 0} \left( \frac{ \alpha({\mathfrak m})^2}{4} \: + \:
\alpha(\sigma)^2 \right), \\
Z_{\Phi}(\sigma, {\mathfrak m}; \tau, {\mathfrak n}) & = & 
\prod_{\rho \in R_{\Phi}} \frac{
\Gamma\left( \frac{R[\Phi]}{2} \: - \: i \rho(\sigma) \: - \:
i f^a[\Phi] \tau_a \: - \: \frac{
\rho({\mathfrak m}) + f^a[\Phi] n_a }{2} \right)
}{
\Gamma\left( 1 \: - \: \frac{R[\Phi]}{2} \: + \: i \rho(\sigma)
\: + \: i f^a[\Phi] \tau_a \: - \: \frac{
\rho({\mathfrak m}) + f^a[\Phi]n_a }{2} \right)
} .
\end{eqnarray*}
The notation above follows \cite{benini1}.
Briefly, $f^a[\Phi]$ encodes the non-R-charges of a chiral
multiplet $\Phi$, and $R[\Phi]$ its R-charge.  $R_{\Phi}$
denotes the corresponding representation of the gauge group.
${\cal W}$ denotes the Weyl group of the gauge group.
$\tau = (\tau_a)$ and ${\mathfrak n}=(n_a)$ define twisted masses for the 
chiral superfield.

The ${\mathfrak m}$ are elements of the Lie algebra of a Cartan subgroup
of the torus, corresponding to elements of the cocharacter or 
dual weight\footnote{ 
The term `dual weight' lattice can be ambiguous:  
some authors occasionally use the term
`weight lattice' to refer to the lattice of weights of 
representations of a particular Lie group, whereas for others
it is defined with respect to the
Lie algebra only, independent of the Lie group.  
The former notion, 
defined in terms of representations of a particular Lie group,
can be more invariantly characterized in 
terms of homomorphisms from the maximal torus of the gauge group to the
circle, and so is often called
the character lattice.  The dual lattice we are interested in here
is similarly known as the cocharacter lattice.
We would like to thank T.~Pantev for a useful discussion of this issue.
}
lattice for the
gauge group, meaning for any representation $R$ of the gauge group
and corresponding
weight $\rho$, $\rho({\mathfrak m}) \in {\mathbb Z}$.  In admittedly
ambiguous notation, we will use ${\mathfrak m}$ to denote both
pertinent Lie algebra elements as well as more abstract elements of the
cocharacter lattice, indexed by tuples of integers.

For our purposes, it will be essential to understand how the sum over
${\mathfrak m}$'s varies depending upon the precise group.  Let us first
compare the sum for $SU(2)$ and $SO(3)$.  Briefly, the character lattice of
$SU(2)$ is twice as large as the character lattice of $SO(3)$, since the
odd-spin representations are only representations of $SU(2)$.  Since
the character lattice of $SU(2)$ is larger, the cocharacter lattice of
$SU(2)$ must be smaller.  In particular, if we normalize such that for
$SO(3)$, ${\mathfrak m}$ varies over all integers, then for $SU(2)$,
${\mathfrak m}$ must vary over only even integers.

Next, let us turn to discrete theta angles in this example.
First, note that ordinary theta angles are encoded in 
$Z_{\rm class}(\sigma, {\mathfrak m})$,
in the
\begin{displaymath}
\exp\left( -i \theta {\rm Tr}\, {\mathfrak m} \right)
\end{displaymath}
factor.  In this notation, in this term we interpret
${\mathfrak m}$ as a matrix in the Lie algebra of a Cartan torus\footnote{
In writing this, we are utilizing the fact that the cocharacter lattice
can be interpreted as a subset of the Cartan torus.
} in
the gauge group, and the trace is taken over that matrix.

Now, for a semisimple gauge group, that trace will always vanish,
and indeed one does not expect ordinary theta angles in two dimensions
in a semisimple gauge theory.  However, that term implicitly tells us
how to insert discrete theta angles in these partition functions ---
by adding a (nonvanishing) term to the (vanishing) theta angle term.
 
In the case of $SO(3)$, a nonzero discrete theta angle
is encoded as a factor
\begin{displaymath}
\exp\left( -i \pi {\mathfrak m} \right) \: = \: (-)^{\mathfrak m}
\end{displaymath}
placed in the same location in the partition function expression as the
ordinary theta angle term.
In other words, for odd ${\mathfrak m}$, corresponding to an 
${\mathfrak m}$ in $SO(3)$'s cocharacter lattice but not $SU(2)$'s, 
this is a sign.
(Note that our notation is ambiguous:  for ordinary theta angles,
we interpret ${\mathfrak m}$ as a matrix, whereas for the discrete
theta angle term, we interpret ${\mathfrak m}$ as an integer.
As both are, morally, the same ${\mathfrak m}$, we will continue
to use ambiguous notation, and trust the reader to disambiguate at
need.)

We can now see, for $SU(2)$ gauge theories with center-invariant
matter, how the decomposition conjecture is realized in partition functions.
Specifically, for $SU(2)$, the partition function~(\ref{eq:part-fn})
is a sum over only even ${\mathfrak m}$; for $SO(3)_+$, a sum over
all integer ${\mathfrak m}$; for $SO(3)_-$, a sum over all
integer ${\mathfrak m}$ but containing an extra $(-)^{\mathfrak m}$.
Schematically, if we write\footnote{
This expression is valid for both the pure supersymmetric theory as
well as a theory with matter:  the difference between the two is encoded
in the function $A({\mathfrak m})$.
We have included an overall factor of (1/2) to reflect
the change in size of the
integration region -- because $SU(2)$ has twice the volume of $SO(3)$,
in principle one would expect that
gauging $SO(3)$ should result in a path integral with a
normalization of $1/2$ relative to gauging an $SU(2)$.  That said, in
quantum field theories not coupled to gravity, overall normalizations of the
partition function are not meaningful, and can be absorbed into counterterms
(see {\it e.g.} \cite{ggk} for a pertinent discussion).  Thus, it might be
better to characterize such choices as conventions.  In that language,
we have chosen convention-dependent factors to make the discussion
more clear.
}
\begin{displaymath}
Z(SO(3)_+) \: = \: \frac{1}{2} 
\sum_{{\mathfrak m}\in {\mathbb Z}} A({\mathfrak m})
\end{displaymath}
for a function $A({\mathfrak m})$ which encodes all the 
${\mathfrak m}$-dependence, then
\begin{displaymath}
Z(SO(3)_-) \: = \: \frac{1}{2}
\sum_{{\mathfrak m}\in {\mathbb Z}} (-)^{\mathfrak m}
A({\mathfrak m})
\end{displaymath}
(for a theory with the same matter),
and 
\begin{displaymath}
Z(SU(2)) \: = \: \sum_{{\mathfrak m} \in 2 {\mathbb Z}} A({\mathfrak m}) \: = \:
Z(SO(3)_+) \: + \: Z(SO(3)_-)
\end{displaymath}
(for an $SU(2)$ theory with the same matter).
Thus, we see that exact $S^2$ partition functions of supersymmetric
gauge $SU(2)$
theories with center-invariant\footnote{
If the matter is not center-invariant, the partition function can still
be written as a sum of two infinite series, but neither infinite series
seems to be interpretable as the partition function of a gauge theory.
In particular, the derivation of $Z_{\Phi}$ for a chiral superfield in
representation $R$ explicitly assumes that $\rho({\mathfrak m})$ is
an integer for all weights $\rho$ and elements ${\mathfrak m}$ of the
cocharacter lattice:  for example, \cite{benini1}[section 3.2] computes
the pertinent operator determinant by expanding in a series of spin spherical
harmonics of spin $(1/2) | \rho({\mathfrak m}) |$, which is only sensible
if $\rho({\mathfrak m})$ is an integer.
More generally, an infinite sum can be written in a variety of ways as
a formal sum of other infinite sums; what is pertinent here is that
the infinite sums have a physical meaning as partition functions of
gauge theories, following the pattern predicted by decomposition.
} 
matter factor in precisely the form predicted by decomposition,
providing a consistency check.

Before describing the general case, let us consider another example:
$SU(3)$.
The simple roots of $SU(3)$ 
can be represented as \cite{georgi}[section 7.2]
\begin{displaymath}
\left( \frac{1}{2}, \frac{\sqrt{3}}{2} \right), \: \: \:
\left( \frac{1}{2}, - \frac{\sqrt{3}}{2} \right)
\end{displaymath}
from which we derive that the cocharacter lattice for
$SU(3)/{\mathbb Z}_3$ has elements of the form
\begin{displaymath}
\left(2 m, \frac{2}{\sqrt{3}} n \right)
\end{displaymath}
for $m, n \in {\mathbb Z}$.  The weights of the fundamental can, in the
same conventions, be represented as
\begin{displaymath}
\left( \frac{1}{2}, \frac{\sqrt{3}}{6} \right), \: \: \:
\left( - \frac{1}{2}, \frac{\sqrt{3}}{6} \right), \: \: \:
\left( 0 , - \frac{\sqrt{3}}{3} \right)
\end{displaymath}
from which we deduce that the cocharacter lattice for $SU(3)$ has
elements of the form
\begin{displaymath}
\left( 2m, \frac{6}{\sqrt{3}} n\right) .
\end{displaymath}
Now, if we are given 
\begin{displaymath}
{\mathfrak m} \: = \: \left(2 m, \frac{2}{\sqrt{3}} n \right)
\end{displaymath}
in the cocharacter lattice of $SU(3)/{\mathbb Z}_3$, the 
${\mathbb Z}_3$-valued
analogue of the second Stiefel-Whiney class is determined by
$n$ mod 3.  Thus, we could add a discrete theta angle to the partition
function of an $SU(3)/{\mathbb Z}_3$ theory by adding a term
\begin{displaymath}
\exp\left( -i \theta n \right)
\end{displaymath}
for $n$ determined by ${\mathfrak m}$ as above and $\theta \in
\{0, 2 \pi/3, 4 \pi/3\}$.  In other words, 
if $w$ denotes the integral of the ${\mathbb Z}_3$-valued
analogue of the Stiefel-Whitney class, then in effect,
\begin{displaymath}
w({\mathfrak m}) \: = \: n \, {\rm mod}\, 3 .
\end{displaymath}

With that in hand, decomposition can be checked for $SU(3)$ theories
in the same form as for $SU(2)$ theories.  As before, we can write
\begin{displaymath}
Z(SU(3)) \: = \: \sum_{m \in {\mathbb Z}} 
\sum_{n \in 3 {\mathbb Z}}A({\mathfrak m})
\end{displaymath}
for the $S^2$ partition function of an $SU(3)$ theory with center-invariant
matter, where ${\mathfrak m}$ is determined by integers $m$, $n$ as above,
and $A({\mathfrak m})$ is defined for ${\mathfrak m}$
corresponding to arbitrary integer $m$, $n$. 
We can decompose this as
\begin{displaymath}
\sum_{m \in {\mathbb Z}} 
\sum_{n \in 3 {\mathbb Z}}A({\mathfrak m})
\: = \: 
\frac{1}{3} \sum_{m \in {\mathbb Z}} 
\sum_{n \in {\mathbb Z}}A({\mathfrak m})
\: + \:
\frac{1}{3} \sum_{m \in {\mathbb Z}} 
\sum_{n \in {\mathbb Z}} e^{- 2\pi i n/3 }
A({\mathfrak m})
\: + \:
\frac{1}{3} \sum_{m \in {\mathbb Z}} 
\sum_{n \in {\mathbb Z}} e^{- 4\pi i n/3 }
A({\mathfrak m}) ,
\end{displaymath}
which can then be interpreted as
\begin{displaymath}
Z(SU(3)) \: = \: Z((SU(3)/{\mathbb Z}_3)_0) \: + \:
Z((SU(3)/{\mathbb Z}_3)_1) \: + \:
Z((SU(3)/{\mathbb Z}_3)_2) ,
\end{displaymath}
precisely in accord with the prediction of decomposition.

It is straightforward to generalize this to more general semisimple
gauge groups $G$.  In general, for semisimple $G$ with center $K$,
if $M_G$ denotes the cocharacter lattice for $G$,
then $M_G \subset M_{G/K}$ and
$M_{G/K}/M_G$ has as many elements as $K$.  In each case, to evaluate
partition functions explicitly, one must represent the integral of
the analogue of the second Stiefel-Whitney class  as an invariant $w$
of the cocharacter lattice, and that invariant determines the pertinent
discrete theta angle term in the partition function.  That invariant $w$
is encoded in the relation\footnote{
See appendix~\ref{ap:cocharacter} for a derivation.
} between the cocharacter lattices:
\begin{displaymath}
1 \: \longrightarrow \: M_G \: \longrightarrow \: 
M_{G/K} \: \stackrel{w}{\longrightarrow} \: K \: \longrightarrow \: 1 .
\end{displaymath}
The decomposition statement is encoded in the fact that
\begin{displaymath} 
\frac{1}{|K|} \sum_{\mu \in \hat{K}} e^{i \mu (w({\mathfrak m}))} 
\end{displaymath}
is a projection operator that projects the lattice $M_{G/K}$ onto
$M_G$, in other words, is the identity on  ${\mathfrak m} \in M_{G/K}$ 
in the image of $M_G$, but vanishes otherwise.

It is now straightforward to check that the partition functions obey
decomposition.  For example, for $G$ semisimple and $K$ a (finite) subgroup
of the center, decomposition in the form~(\ref{eq:decomp-simple2}) can
be checked as follows:
\begin{eqnarray*}
Z(G) & = & \sum_{{\mathfrak m} \in M_G} A({\mathfrak m}), \\
& = &
\frac{1}{|K|}
\sum_{\lambda \in \hat{K}}
\sum_{{\mathfrak m} \in M_{G/K}} e^{i \lambda( w({\mathfrak m})) } 
A({\mathfrak m}),
\\
& = & \sum_{\lambda \in \hat{K}}
Z( (G/K)_{\lambda} ),
\end{eqnarray*}
where $M_G$ is the cocharacter lattice of $G$, and
\begin{displaymath}
Z( (G/K)_{\lambda} ) \: = \:
\frac{1}{|K|}
\sum_{{\mathfrak m} \in M_{G/K}} e^{i \lambda( w({\mathfrak m})) } 
A({\mathfrak m}).
\end{displaymath}

The same methods can also be applied to check decomposition in the
more general form~(\ref{eq:decomp-general}):
for $G = \tilde{G}/K$, $G$ semisimple, $\tilde{G}$ simply-connected,
$K$ a subgroup of the center of $\tilde{G}$,
and $\lambda \in \hat{K}$,
\begin{eqnarray*}
Z( (G)_{\lambda} ) & = &
\sum_{{\mathfrak m} \in M_G} e^{i \lambda( w_K({\mathfrak m}))}
A({\mathfrak m}), \\
& = & 
\frac{1}{|C(G)|}
\sum_{\mu \in \widehat{C(G)} }
\sum_{{\mathfrak m} \in M_{G/C(G)}} 
e^{i (\lambda,\mu)  w_{C(\tilde{G})}({\mathfrak m}) }
A({\mathfrak m}),
\\
& = &
\sum_{\mu \in \widehat{C(G)} }
Z( (G/C(G))_{\lambda, \mu} ).
\end{eqnarray*}
In the expression above, $w_K: M_G \rightarrow K$ (with kernel $M_{\tilde{G}}$),
$w_{C(\tilde{G})}: M_{G/C(G)} (= M_{\tilde{G}/C(\tilde{G})})
 \rightarrow C(\tilde{G})$ with kernel $M_{\tilde{G}}$,
and $w_{C(G)}: M_{G/C(G)} \rightarrow C(G)$ with kernel $M_G$.
The projection operator has been modified slightly:  we have replaced
the factor
\begin{displaymath}
e^{i \mu( w_{C(G)}({\mathfrak m}))}
\end{displaymath}
(which reduces to $1$ on the image of $M_G$ in $M_{G/C(G)}$)
with
\begin{displaymath}
e^{i (\lambda,\mu)  w_{C(\tilde{G})}({\mathfrak m}) }.
\end{displaymath}
We can see that this reduces to the desired expression on $M_G$ as follows.
Because\footnote{
See appendix~\ref{ap:cocharacter} for a derivation.
} 
$w_{C(G)} = \alpha \circ w_{C(\tilde{G})}$ where
\begin{displaymath}
1 \: \longrightarrow \: K \: \longrightarrow \: C(\tilde{G}) \: 
\stackrel{\alpha}{\longrightarrow} \: C(G) \: \longrightarrow \: 1,
\end{displaymath}
and the fact that
the image of $M_G$ is in the kernel of $w_{C(G)}$, we see that the
image of $M_G$ in $M_{\tilde{G}/C(\tilde{G})} = M_{G/C(G)}$ must lie in
the image of $K$, hence from equation~(\ref{eq:prod-defn}) and
commutivity\footnote{
See appendix~\ref{ap:cocharacter} for a 
derivation.}
of the square
\begin{displaymath}
\xymatrix{
K \ar[r] & C(\tilde{G}) \\
M_{G=\tilde{G}/K} \ar[u]^{w_K}  \ar[r] & M_{\tilde{G}/C(\tilde{G})} 
\ar[u]^{w_{C(\tilde{G})}} 
}
\end{displaymath}
we find that
\begin{displaymath}
e^{i (\lambda,\mu)  w_{C(\tilde{G})}({\mathfrak m}) } \: = \:
e^{i \lambda w_{K}({\mathfrak m}) }
\end{displaymath}
for ${\mathfrak m}$ in the image of $M_G$.  Thus, our modified projection
operator reduces to the desired form on $M_G$.

\section{Four dimensional theories}  \label{sect:4d}

In four dimensional theories, it is no longer the case that a gauge
theory with a trivially-acting subgroup is physically equivalent
to a gauge theory with a restriction on instantons.  For example,
in four dimensions, pure $N=1$ supersymmetric $SU(2)$ Yang-Mills 
theory does not violate cluster decomposition, and has the same number of
vacua in the IR as pure $N=1$ supersymmetric $SO(3)$ Yang-Mills theory.

However, it is nonetheless intriguing to compare, for example, $SU(2)$ gauge
theories to $SO(3)$ gauge theories with restrictions on instantons.

In this section, we will discuss four-dimensional gauge theories with
restrictions on instantons, enforced by inserting projection operators
into path integrals.  These theories do not obey cluster decomposition,
but, at least for SCFT's, have been argued 
to obey an analogue of the decomposition
conjecture \cite{hs}[appendix A], 
so that cluster decomposition is violated in an extremely
mild way.  That said, unlike decomposition in two dimensions, there has
been very little work done on decomposition in four dimensions,
so part of the purpose of this section is to
further illuminate decomposition in four dimensional theories.
For simplicity, we will restrict to four-dimensional SCFT's, both because these
were the theories considered in \cite{hs}[appendix A], and also so that we can
reliably work in a weak coupling regime.

We will first examine dyon spectra in four-dimensional
theories with restrictions on
instanton sectors.  Reducing the number of instantons changes the theta
angle periodicity, hence one can in principle run into a consistency
problem in 
theta angle charge rotations \cite{witten-dyons}.  If the theory
decomposes into a disjoint union of (ordinary) theories, a contradiction is
averted.
Furthermore,
we will see that dyon spectra can be used to distinguish an $SU(2)$
gauge theory from an $SO(3)$ gauge theory with a restriction on instantons
obeying an analogue of decomposition.
We will also briefly
examine Vafa-Witten topological field theory partition functions, as they can 
be used to provide simple examples of the four dimensional decomposition
of \cite{hs}[appendix A] in action.

\subsection{Dyon charge lattices}

Let us begin by examining dyon spectra in four dimensional theories
with restrictions on instantons.  Start with a
four-dimensional gauge theory with conventions chosen such that
$\theta$ is $2\pi$-periodic.  Restrict allowed instantons to those with
instanton numbers divisible by some integer $k>0$.  The restricted
theory has a different $\theta$ periodicity:  the theory is invariant
under $\theta \equiv \theta + 2 \pi/k$.

Because of how dyon charges depend upon the $\theta$ angle \cite{witten-dyons},
this constrains dyon spectra.  In the original
theory, the physics and so the dyon spectrum\footnote{
Individual dyon charges rotate as described in \cite{witten-dyons}, but the
spectrum as a whole remains invariant.
} was invariant under
$\theta \mapsto 
\theta + 2 \pi$, but in the new theory, it must be invariant under
$\theta \mapsto \theta + 2 \pi/k$.

This would appear to pose a consistency problem for the theory.
If, for example, we have a dyon of charge $(\lambda_e, \lambda_m)$,
then under a $2 \pi/k$ rotation of $\theta$, the dyon charge would
become\footnote{
The dyon charge formula, although it depends upon $\theta$, is a purely
semiclassical result independent of instanton sector \cite{witten-dyons},
and so it is not affected by the restriction on allowed instantons.
}
\cite{witten-dyons}
$(\lambda_e + \lambda_m/k, \lambda_m)$, which, except for special values of
$k$, is unlikely to land somewhere in the original charge lattice.

Nevertheless, it is still possible for the charge lattice to be closed
under such a rotation.  One
universal solution is if the dyon spectrum is a sum of copies of the dyon
spectrum of the original theory, but with $\theta$ rotated in increments
of $2 \pi/k$, as arises in a disjoint union of theories on the
same spacetime.  (This is a prediction of the conjectured analogue
of the decomposition conjecture
of \cite{summ} in four-dimensional theories \cite{hs}.)  Ordinarily locality
as discussed in \cite{ast}[section 1.1] would prohibit dyon spectra of
this form; however,
if we have a disjoint union of theories, so that dyons associated with
different theories do not interact, then locality only applies within each
separate sector, and so any contradiction is averted.

The solution above is not unique, and we will see that in special cases,
other solutions are possible.  One example we will study in detail
will involve the relation between an $SO(3)$ gauge theory with a
restriction on instantons, and an $SU(2)$ theory.  We will see that
the $SU(2)$ theory encodes a different solution to the corresponding
problem in this context.

Let us consider a concrete example to make this proposal more precise.
Consider an $SU(2)$ gauge theory with a restriction on instantons,
restricted to instantons with instanton number divisible by 3.
The dyon spectrum of the original $SU(2)$ theory itself has the form
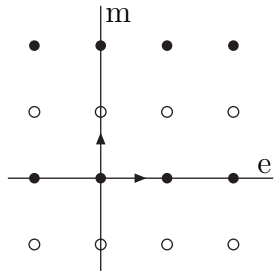
\cite{ast}
\begin{center}
\begin{picture}(100,100)
\ArrowLine(0,35)(100,35)
\ArrowLine(35,0)(35,100)
\Text(97,37)[b]{e}
\Text(37,97)[l]{m}
\Vertex(35,35){2}  \Vertex(85,35){2}
\Vertex(35,85){2}  \Vertex(85,85){2}
\CArc(10,10)(2,0,360) \CArc(35,10)(2,0,360) \CArc(60,10)(2,0,360)
\CArc(85,10)(2,0,360)
\CArc(10,60)(2,0,360) \CArc(35,60)(2,0,360)  \CArc(60,60)(2,0,360)
\CArc(85,60)(2,0,360)
\Vertex(10,35){2} \Vertex(60,35){2}
\Vertex(10,85){2} \Vertex(60,85){2}
\end{picture}
\end{center}
(Allowed electric charges are arbitrary integers; allowed magnetic charges
are even integers.)
Under $\theta \mapsto \theta + 2 \pi/3$, the charges rotate
as $(\lambda_e,\lambda_m) \mapsto (\lambda_e + \lambda_m/3, \lambda_m)$,
and the resulting charge lattice has the form
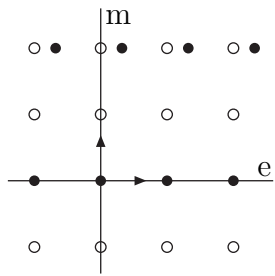
\begin{center}
\begin{picture}(100,100)
\ArrowLine(0,35)(100,35)
\ArrowLine(35,0)(35,100)
\Text(97,37)[b]{e}
\Text(37,97)[l]{m}
\Vertex(35,35){2}  \Vertex(85,35){2}
\CArc(35,85)(2,0,360)  \CArc(85,85)(2,0,360)
\Vertex(43,85){2}  \Vertex(93,85){2}
\CArc(10,10)(2,0,360) \CArc(35,10)(2,0,360) \CArc(60,10)(2,0,360)
\CArc(85,10)(2,0,360)
\CArc(10,60)(2,0,360) \CArc(35,60)(2,0,360)  \CArc(60,60)(2,0,360)
\CArc(85,60)(2,0,360)
\Vertex(10,35){2} \Vertex(60,35){2}
\CArc(10,85)(2,0,360)  \CArc(60,85)(2,0,360)
\Vertex(18,85){2}  \Vertex(68,85){2}
\end{picture}
\end{center}
Here, the electric charges of dyons with $\lambda_m=2$ are shifted by $1/3$.
Similarly, the electric charges of dyons with $\lambda_m=4$ would be shifted
by $2/3$, and so forth.

As anticipated above, under this rotation, the dyon spectrum is not invariant.
However, the physics should be invariant, as we are rotating the theta
angle by its periodicity.  As suggested above, one solution is that the
correct dyon spectrum is determined by a set of dyons arising from variable
$\theta$ angles.  The resulting dyon spectrum is given by the sum of the two
spectra above, plus another resulting from rotating $\theta \mapsto
\theta + 4 \pi/3$, to get
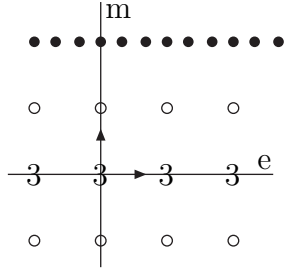
\begin{center}
\begin{picture}(100,100)
\ArrowLine(0,35)(100,35)
\ArrowLine(35,0)(35,100)
\Text(97,37)[b]{e}
\Text(37,97)[l]{m}
\Text(32,35)[l]{3}  \Text(82,35)[l]{3}
\Vertex(35,85){2}  \Vertex(85,85){2}
\Vertex(43,85){2}  \Vertex(93,85){2}
\Vertex(52,85){2}  \Vertex(102,85){2}
\CArc(10,10)(2,0,360) \CArc(35,10)(2,0,360) \CArc(60,10)(2,0,360)
\CArc(85,10)(2,0,360)
\CArc(10,60)(2,0,360) \CArc(35,60)(2,0,360)  \CArc(60,60)(2,0,360)
\CArc(85,60)(2,0,360)
\Text(7,35)[l]{3}  \Text(57,35)[l]{3}
\Vertex(10,85){2} \Vertex(60,85){2}
\Vertex(18,85){2}  \Vertex(68,85){2}
\Vertex(27,85){2}  \Vertex(77,85){2}
\end{picture}
\end{center}
We have used ``$3$'''s to indicate that there are three separate dyons
with the same listed charges.
As noted earlier, this dyon spectrum is only sensible if the theory
decomposes into a disjoint union; our point is that such a decomposition
gives a dyon spectrum compatible with theta angle dyon charge rotations.

Another example will be instructive.  Let us compare $SU(2)$ to the
$SO(3)_{\pm}$
gauge theories.  These are very similar:  all have the same perturbative
description, for example, as the Lie algebras match.  Furthermore, the
$SU(2)$ instantons are a subset of the $SO(3)$ instantons, which suggests
(incorrectly) that the $SU(2)$ theory is equivalent to an $SO(3)$ theory
with a restriction on instantons.  Let us work through this in detail,
to see how the $SU(2)$ theory is distinguished from $SO(3)_{\pm}$ with
a restriction on instantons.

Specifically,
consider the four-dimensional $SO(3)_-$ theory \cite{ast,gmn}.
Let us restrict the instantons of the $SO(3)_-$ theory to
those with instanton numbers divisible by 2 (in conventions in which
the instantons of $SO(3)_-$ have integral instanton number).
The original $SO(3)_-$ theory is periodic under $\theta \mapsto \theta
+ 4 \pi$; the new theory with the restriction on instanton numbers
is periodic under $\theta \mapsto \theta + 2 \pi$.

The prediction above is that in this restricted theory, the dyon spectrum
is the sum of the dyons in two different theories:
\begin{displaymath}
SO(3)_-^{\theta} \mbox{ and }
SO(3)_-^{\theta + 2 \pi} \: = \: SO(3)_+^{\theta} .
\end{displaymath}
(Hand-in-hand, the physical theory must similarly decompose into a 
disjoint union.)
In the figure below, we illustrate dyon charge multiplicities occuring
in the resulting spectrum:
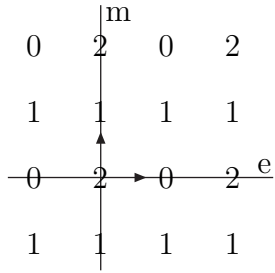
\begin{center}
\begin{picture}(100,100)
\ArrowLine(0,35)(100,35)
\ArrowLine(35,0)(35,100)
\Text(97,37)[b]{e}
\Text(37,97)[l]{m}
\Text(32,35)[l]{2}  \Text(82,35)[l]{2}
\Text(32,85)[l]{2}  \Text(82,85)[l]{2}
\Text(7,10)[l]{1} \Text(32,10)[l]{1} \Text(57,10)[l]{1} \Text(82,10)[l]{1}
\Text(7,60)[l]{1} \Text(32,60)[l]{1} \Text(57,60)[l]{1} \Text(82,60)[l]{1}
\Text(7,35)[l]{0} \Text(57,35)[l]{0}
\Text(7,85)[l]{0} \Text(57,85)[l]{0}
\end{picture}
\end{center}
As a consistency check, under the rotation $\theta \mapsto \theta + 2 \pi$,
note that the lattice as a whole is invariant under
the dyon charge rotation
\cite{witten-dyons}
\begin{displaymath}
\left( \lambda_e, \lambda_m \right) \: \mapsto \:
\left( \lambda_e + \lambda_m, \lambda_m \right) .
\end{displaymath}
For example, when $\lambda_m$ is even, the only allowed $\lambda_e$'s are
all even, and so they shift into one another.  When $\lambda_m$ is odd,
all $\lambda_e$'s are permitted, and they shift into one another.

In this case, the solution above is not unique.  Another dyon spectrum
which is also closed under $\theta \mapsto \theta + 2\pi$ is of the form
\begin{center}
\begin{picture}(100,100)
\ArrowLine(0,35)(100,35)
\ArrowLine(35,0)(35,100)
\Text(97,37)[b]{e}
\Text(37,97)[l]{m}
\Text(32,35)[l]{1}  \Text(82,35)[l]{1}
\Text(32,85)[l]{1}  \Text(82,85)[l]{1}
\Text(7,10)[l]{0} \Text(32,10)[l]{0} \Text(57,10)[l]{0} \Text(82,10)[l]{0}
\Text(7,60)[l]{0} \Text(32,60)[l]{0} \Text(57,60)[l]{0} \Text(82,60)[l]{0}
\Text(7,35)[l]{1} \Text(57,35)[l]{1}
\Text(7,85)[l]{1} \Text(57,85)[l]{1}
\end{picture}
\end{center}
and is significantly different from the spectrum above.
For example, in this spectrum we do not have multiple copies of any
dyons; this theory appears to be ``connected,'' loosely speaking.
In fact,
this is precisely the spectrum of the $SU(2)$ theory \cite{ast}.
Other than the spectrum of dyons, however,
the restricted $SO(3)_-$ and $SU(2)$ theories are very
similar:  both have the same perturbative description, both have the
same instantons, but the dyon spectrum in the two cases is significantly
different.

\subsection{Vafa-Witten TFT partition functions}

As four dimensional decomposition has not been checked nearly as thoroughly
as two dimensional cases, it is useful to study further examples.
To that end,
in this section we will discuss partition functions of the
topological field theories of \cite{vw},
arising from topologically-twisted four-dimensional $N=4$ theories, 
with a restriction
on instantons.
We shall see that those partition functions can be written as 
partition functions of disjoint unions of theories,
in agreement the four-dimensional
decomposition prediction of \cite{hs}[appendix A].
In fact, the same arguments can be applied verbatim to other supersymmetric
instanton computations, such as the Nekrasov partition function, as the
partition functions have the same general form.  For simplicity, we will
only describe Vafa-Witten partition functions.
(Of course, a thorough check of decomposition in this case
would require more than
comparing partition functions -- we would need to also demonstrate that
correlation functions factorize, for example, as has been shown in
related two-dimensional theories. Our purpose here is merely to provide some
evidence supporting the argument of \cite{hs}[appendix A].)

Briefly, the partition functions of these topological field
theories are of the form \cite{vw}
\begin{displaymath}
Z(q) \: = \: \sum_n a_n q^n ,
\end{displaymath}
where
\begin{displaymath}
q \: = \: e^{2 \pi i \tau} \: = \:
\exp\left( i \theta \: - \: \frac{8 \pi^2}{g^2} \right) ,
\end{displaymath}
and $a_k$ is proportional to the moduli space of
$n$ instantons.  Suppose, for example, that we consider the
analogous theory defined by a restriction to instanton numbers divisible
by $k$.  Then the partition function of this new theory would be
\begin{displaymath}
Z_k(q) \: = \: \sum_n a_{nk} q^{nk} .
\end{displaymath}
Note that this is the same as
\begin{displaymath}
\sum_n \left( \frac{1}{k} \sum_{\alpha=0}^{k-1} e^{2 \pi i \alpha n / k} 
\right)
a_n q^n \: = \:
\frac{1}{k} \sum_{\alpha = 0}^{k-1} 
\sum_n a_n q^n e^{2 \pi i \alpha n / k} .
\end{displaymath}
The effect of the insertion is to project onto instanton numbers divisible
by $k$:  when $n$ is divisible by $k$, each factor in the $\alpha$ sum
is $1$, otherwise the $\alpha$ sum is a sum over roots of unity, which
vanishes.  At the same time, writing the partition function for the theory
with a restriction on instanton sectors, in this form,
\begin{eqnarray*}
Z_k(q) & = & \frac{1}{k} \sum_{\alpha = 0}^{k-1} 
\sum_n a_n q^n e^{2 \pi i \alpha n / k}
\: = \: \frac{1}{k} \sum_{\alpha = 0}^{k-1} 
\sum_n a_n \exp \left( n\left(
i \theta \: + \: i 2 \pi \alpha/k \: - \: \frac{8 \pi^2}{g^2} \right) \right),
\\
& = & \frac{1}{k} \sum_{\alpha=0}^{k-1}
Z(q_{\alpha}),
\end{eqnarray*}
where
\begin{displaymath}
q_{\alpha} \: = \: 
 \exp\left( i \theta \: + \: i 2 \pi \alpha/k \: - \: \frac{8 \pi^2}{g^2} 
\right), 
\end{displaymath}
consistent with an interpretation as a disjoint union of theories with rotating
$\theta$ angles, as predicted in \cite{hs}.

\section{Dijkgraaf-Witten theory}  \label{sect:dw}

The recent paper
\cite{kaps} studied strings on gerbes, the same sorts of
theories we have discussed in this paper, albeit described there
in the language of
QFT's coupled to TQFT's.  The prototype for the TQFT coupling in
\cite{kaps} was given by
Dijkgraaf-Witten theory \cite{dw1}.  Briefly, we can understand
analogues of
Dijkgraaf-Witten theory in $n$ dimensions, as defined by \cite{kaps},
directly in finite gauge theory,
without the overhead of various abelian gauge theories.
In such language, the theories of \cite{kaps} can be described 
as an $n$-dimensional orbifold
of a point with an analogue of discrete torsion turned on.

For example, in two dimensions with gauge group $G$, the partition function
of the two-dimensional version\footnote{
The original reference \cite{dw1} only considered three-dimensional theories,
but glossing over occasional subtleties such as certain functoriality
issues in low dimensions, analogues are straightforward to write down in
any dimension, as indeed \cite{kaps} do.
} of Dijkgraaf-Witten theory on $T^2$ is
given by
\begin{displaymath}
Z_2 \: = \: \frac{1}{|G|} \sum_{gh = hg} \frac{ \alpha(g,h) }{ \alpha(h,g) },
\end{displaymath}
where $\alpha \in H^2(G,U(1))$ is a group cocycle (with trivial action
on the coefficients).  The ratio
\begin{displaymath}
\frac{ \alpha(g,h) }{ \alpha(h,g) }
\end{displaymath}
is invariant under both coboundaries and $SL(2,{\mathbb Z})$.

In three dimensions, with gauge group $G$ the partition function of
Dijkgraaf-Witten theory has a similar form, with phase factors now given
by the analogues of discrete torsion for the $C$ field \cite{cdt,dtrev}.
For example, on a three-torus, the partition function is given by
\begin{displaymath}
Z_3 \: = \: \frac{1}{|G|} \sum_{{\rm commuting}\:{\rm triples}}
\prod_{{\rm permutations}\:P} \alpha(g_{i_1}, g_{i_2}, g_{i_3})^{\pm P} .
\end{displaymath}
The alternating ratio of cocycles $\alpha \in H^3(G,U(1))$ can be shown to
be both invariant under coboundaries and also invariant under
$SL(3,{\mathbb Z})$ \cite{dw1,cdt}.

Similarly, in $d$ dimensions, for gauge group $G$ the analogue of
Dijkgraaf-Witten theory has a similar form.  For example, on a $d$-torus,
the partition function is given by
\begin{displaymath}
Z_d \: = \: \frac{1}{|G|} \sum_{{\rm commuting}\:{\rm pairs}}
\prod_{{\rm permutations}\:P} \alpha(g_{i_1}, g_{i_2}, \cdots,
g_{i_d})^{\pm P} ,
\end{displaymath}
where $\alpha$ is a cocycle representing an element of $H^d(G,U(1))$, with
trivial action on the coefficients.  (It is straightforward to check
that this expression is invariant under coboundaries.)
(See also \cite{aspinwall-dt} for a derivation of such expressions
from simplices in two dimensions, and \cite{dw1} for an analogous derivation
in three dimensions.)

Now, turning on discrete torsion in a trivially-acting subgroup of
a gauge group in two dimensions modifies the decomposition conjecture,
as observed in \cite{summ}[section 10].  In particular, in the presence
of discrete torsion, decomposition may no longer take place at all.

A prototypical example is provided by the two-dimensional orbifold
$[X / ({\mathbb Z}_2 \times {\mathbb Z}_2)]$, where $X$ is a manifold
(defining a nonlinear sigma model) and the first
${\mathbb Z}_2$ acts trivially on $X$. 
It was argued in \cite{summ}[section 10.1] 
that in this case,
if discrete torsion is turned on,
the theory on $[X / ({\mathbb Z}_2 \times {\mathbb Z}_2)]$ is equivalent
to a sigma model on one copy of $X$ -- a 2-fold cover of $[X/{\mathbb Z}_2]$.  
More generally, for the two-dimensional
orbifold $[X / ({\mathbb Z}_k \times {\mathbb Z}_k)]$ where the
first ${\mathbb Z}_k$ acts trivially, for any nontrivial discrete torsion,
this theory is equivalent\footnote{
In passing, we should mention that there is also related work in the
math literature, albeit describing a different duality.  Specifically,
the paper \cite{stt} gives a mathematical proposal for understanding
decomposition with discrete torsion turned on in the fashion indicated
here.  However, in this example, the duality described in \cite{stt}
predicts \cite{tsengpriv} $k$ copies of $[X/{\mathbb Z}_k]$ rather
than one copy of $X$.   
} \cite{summ}[section 10.2]
to a sigma model on one copy of $X$, a $k$-fold
cover of $[X/{\mathbb Z}_k]$.  (The fastest way to see this is to compute
the genus one worldsheet partition function -- because of the
discrete torsion phase factors, all contributions from twisted sectors
cancel out, leaving just the partition function of an ordinary nonlinear
sigma model on $X$.)

\section{Conclusions}

In this paper we have discussed `decomposition' in two and four dimensions.
In two dimensions, it is a long established result that in orbifolds and
abelian gauge theories, if a finite subgroup leaves matter invariant, then
the theory `decomposes' into a disjoint union of theories.  As there is
no gauge dynamics in two dimensions, one therefore expects closely related
phenomena in two-dimensional nonabelian gauge theories.  We gave a general
conjecture for its formulation, and demonstrated that both nonsupersymmetric
pure two-dimensional Yang-Mills and supersymmetric theories in two dimensions
obey the general principle.  In particular, this allowed us to derive
a decomposition result for pure two-dimensional nonsupersymmetric
Yang-Mills.

In four dimensions, existence of a trivially-acting finite subgroup
is no longer equivalent to a restriction on instantons.
We discussed how these can be distinguished using dyon spectra, and
also discussed partition function constructions in four-dimensional
topological field theories with restrictions on instantons that
mirror analogous two-dimensional constructions.

There are a number of other directions to pursue.  For one example,
it is well-known that two-dimensional Yang-Mills can also be formulated as a
string theory (see {\it e.g.} \cite{cmr-str,cmr} 
for a complete description of that
string theory).  It would be interesting to understand how that string
theory also decomposes.

For another example, zero-area limits of partition functions of 
$q$-deformed pure two-dimensional Yang-Mills are now understood to compute
some indices of four-dimensional superconformal field theories,
see {\it e.g.} \cite{grry1,grry2}.  It would be interesting to understand
the implications of the decomposition conjecture described here for
four-dimensional superconformal indices.  Perhaps, although full
four dimensional
theories do not obey the same decomposition (unless one enforces a 
strong instanton restriction), the indices decompose.

Yet another direction to pursue is decomposition in three dimensions.
Do three-dimensional theories with trivially-acting finite groups 
decompose, or only if one imposes a further restriction on instantons?
Such questions would be interesting to understand.

\section{Acknowledgements}

We would like to thank R.~Donagi, S.~Hellerman, B.~Jia, S.~Katz,
T.~Pantev, Y.~Tachikawa, X.~Tang, and H.-H.~Tseng for useful
discussions.
E.S. was partially supported by NSF grant PHY-1068725.

\appendix

\section{Miscellaneous cocharacter lattice relations}
\label{ap:cocharacter}

In this appendix we shall derive a few relations between cocharacter
lattices that are used in the text.

First, for semisimple $G$ with $K$ a finite subgroup of the center, if $M_G$
denotes the cocharacter lattice for $G$, then we shall show that
\begin{displaymath}
1 \: \longrightarrow \: M_G \: \longrightarrow \: 
M_{G/K} \: \stackrel{w}{\longrightarrow} \: K \: \longrightarrow \: 1 .
\end{displaymath}

This relation can be derived as follows \cite{tonypriv}.
Begin with the short exact
sequence
\begin{displaymath} 
1 \: \longrightarrow \: K \: \longrightarrow \: T_G \: \longrightarrow
\: T_{G/K} \: \longrightarrow \: 1,
\end{displaymath}
and apply the functor Hom$(U(1),-)$.  The result is
\begin{displaymath}
1 \: \longrightarrow \: M_G \: \longrightarrow \: M_{G/K} \: \longrightarrow
\: {\rm Ext}^1(U(1),K) \: \longrightarrow \: 1.
\end{displaymath}
To compute Ext$^1$ above, apply Hom$(-,K)$ to
\begin{displaymath}
0 \: \longrightarrow \: {\mathbb Z} \: \longrightarrow \:
{\mathbb R} \: \longrightarrow \: U(1) \: \longrightarrow \: 1
\end{displaymath}
to get
\begin{displaymath}
1 \: \longrightarrow \: {\rm Hom}(U(1),K) \: \longrightarrow \:
{\rm Hom}({\mathbb R},K) \: \longrightarrow \:
{\rm Hom}({\mathbb Z},K) \: \longrightarrow \:
{\rm Ext}^1(U(1),K) \: \longrightarrow \: {\rm Ext}^1({\mathbb R},K).
\end{displaymath}
Now,
\begin{displaymath}
{\rm Hom}(U(1),K) \: = \: 1 \: = \: {\rm Hom}({\mathbb R},K)
\end{displaymath}
since $U(1)$, ${\mathbb R}$ are divisible groups and $K$ is torsion, and
similarly Ext$^1({\mathbb R},K)=0$ since ${\mathbb R}$ is torsion-free and $K$
is torsion.  Thus,
\begin{displaymath}
{\rm Ext}^1(U(1),K) \: \cong \: {\rm Hom}({\mathbb Z},K) \: = \: K
\end{displaymath}
instead of $\hat{K}$.
We would like to thank T.~Pantev for sharing this computation.
We will denote the map $w: M_{G/K} \rightarrow K$ by $w_K$.

Next, define the map $\alpha: C(\tilde{G}) \rightarrow C(G)$,
where $C(G)$ denotes the center of $G$, as the second nontrivial
map in the sequence
\begin{displaymath}
1 \: \longrightarrow \: K \: \longrightarrow \: C(\tilde{G}) \: 
\stackrel{\alpha}{\longrightarrow} \: C(G) \: \longrightarrow \: 1.
\end{displaymath}
We shall show that the following diagram commutes:
\begin{displaymath}
\xymatrix{
K \ar[r] & C(\tilde{G}) \ar[r]^{\alpha} & C(G) \\
M_{G=\tilde{G}/K} \ar[u]^{w_K}  \ar[r] & M_{\tilde{G}/C(\tilde{G})} 
\ar[u]^{w_{C(\tilde{G})}} 
\ar@{=}[r] & M_{G/C(G)} \ar[u]^{w_{C(G)}}
}
\end{displaymath}
To show this, begin with the diagram
\begin{displaymath}
\xymatrix{
1 & 1 & 1 \\
T_{G=\tilde{G}/K} \ar[u] \ar[r] & T_{\tilde{G}/C(\tilde{G})} \ar[u] 
\ar@{=}[r] & T_{G/C(G)} \ar[u] 
\\
T_{\tilde{G}} \ar[u] \ar@{=}[r] & T_{\tilde{G}} \ar[u] \ar[r] &
T_{G = \tilde{G}/K} \ar[u] \\
K \ar[u] \ar[r] & C(\tilde{G}) \ar[u] \ar[r]^{\alpha} & C(G) \ar[u] \\
1 \ar[u] & 1 \ar[u] & 1 \ar[u]
}
\end{displaymath}
This commutes, hence the diagram obtained by applying Hom$(U(1),-)$,
namely
\begin{displaymath}
\xymatrix{
{\rm Ext}^1 (U(1),K) \ar[r] & {\rm Ext}^1(U(1),C(\tilde{G}) \ar[r]^{\alpha} &
{\rm Ext}^1(U(1), C(G)) \\
M_{G=\tilde{G}/K} \ar[u]^{w_K}  \ar[r] & M_{\tilde{G}/C(\tilde{G})} 
\ar[u]^{w_{C(\tilde{G})}} 
\ar@{=}[r] & M_{G/C(G)} \ar[u]^{w_{C(G)}} \\
M_{\tilde{G}} \ar[u] & M_{\tilde{G}} \ar[u] & M_G \ar[u]
}
\end{displaymath}
also commutes, which is the desired result.


\begin{thebibliography}{199}

\addcontentsline{toc}{section}{References}

\bibitem{summ} S. Hellerman, A. Henriques, T. Pantev, E. Sharpe, M. Ando,
``Cluster decomposition, T-duality, and gerby CFT's,''
Adv. Theor. Math. Phys. {\bf 11} (2007) 751-818,
{\tt hep-th/0606034}.

\bibitem{ggp1} A. Gadde, S. Gukov, P. Putrov,
``(0,2) trialities,'' JHEP 1403 (2014) 076,
{\tt arXiv:  1310.0818}.

\bibitem{kl1} D. Kutasov, J. Lin, ``(0,2) dynamics from four dimensions,''
{\tt arXiv:  1310.6032}.

\bibitem{jsw} B. Jia, E. Sharpe, R. Wu, ``Notes on nonabelian (0,2) theories
and dualities,'' {\tt arXiv:  1401.1511}.

\bibitem{kl2} D. Kutasov, J. Lin, ``(0,2) ADE models from four dimensions,''
{\tt arXiv:  1401.5558}.

\bibitem{bpv} N. Bobev, K. Pilch, O. Vasilakis, 
``(0,2) SCFT's from the Leigh-Strassler fixed point,''
{\tt arXiv:  1403.7131}.


\bibitem{nr} T. Pantev, E. Sharpe, ``Notes on gauging noneffective
group actions,'' {\tt hep-th/0502027}.

\bibitem{msx} T. Pantev, E. Sharpe, ``String compactifications on
Calabi-Yau stacks,'' Nucl. Phys. {\bf B733} (2006) 233-296,
{\tt hep-th/0502044}.

\bibitem{glsm} T. Pantev, E. Sharpe, ``GLSM's for gerbes (and other toric
stacks),'' Adv. Theor. Math. Phys. {\bf 10} (2006) 77-121,
{\tt hep-th/0502053}.

\bibitem{cdhps} A. Caldararu, J. Distler, S. Hellerman, T. Pantev,
E. Sharpe, ``Non-birational twisted derived equivalences in abelian GLSMs,''
Comm. Math. Phys. {\bf 294} (2010) 605-645,
{\tt arXiv:  0709.3855}.


\bibitem{hs} S. Hellerman, E. Sharpe, ``Sums over topological sectors
and quantization of Fayet-Iliopoulos parameters,''
Adv. Theor. Math. Phys. {\bf 15} (2011) 1141-1199,
{\tt arXiv:  1012.5999}.

\bibitem{bgmru} M. Berasaluce-Gonz\'alez, M. Montero, A. Retolaza,
A. M. Uranga, ``Discrete gauge symmetries from (closed string) tachyon
condensation,'' {\tt arXiv:  1305.6788}.

\bibitem{addss} N. Addington, E. Segal, E. Sharpe, ``D-brane probes,
branched double covers, and noncommutative resolutions,''
{\tt arXiv:  1211.2446}.

\bibitem{hkm} J. Halverson, V. Kumar, D. Morrison, ``New methods for
characterizing phases of 2d supersymmetric gauge theories,''
{\tt arXiv:  1305.3278}.

\bibitem{es-rflat} E. Sharpe, ``A few Ricci-flat stacks as phases of
exotic GLSM's,'' Phys. Lett. {\bf B726} (2013) 390-395,
{\tt arXiv:  1306.5440}.

\bibitem{hori2} K. Hori, ``Duality in two-dimensional (2,2) supersymmetric
nonabelian gauge theories,'' JHEP 1310 (2013) 121,
{\tt arXiv:  1104.2853}.

\bibitem{hetgerbe} L. Anderson, B. Jia, R. Manion, B. Ovrut, E. Sharpe,
``General aspects of heterotic string compactifications on stacks and
gerbes,'' {\tt arXiv:  1307.2269}.

\bibitem{me-vienna} E. Sharpe, ``Derived categories and stacks in
physics,'' contribution to the proceedings of the ESI research conference
on homological mirror symmetry (Vienna, Austria, June 2006), 
{\tt arXiv:  hep-th/0608056}.

\bibitem{me-tex} E. Sharpe, ``Landau-Ginzburg models, gerbes, and
Kuznetsov's homological projective duality,'' to appear in the proceedings
of {\it Topology, ${\bf C}^*$ algebras, string duality} (Texas Christian
University, May 18-22, 2009).

\bibitem{me-qts} E. Sharpe, ``GLSM's, gerbes, and Kuznetsov's homological
projective duality,'' contribution to the proceedings of
{\it Quantum theory and symmetries 6}, {\tt arXiv:  1004.5388}. 

\bibitem{ajt1} E. Andreini, Y. Jiang, H.-H. Tseng, ``On Gromov-Witten theory
of root gerbes,'' {\tt arXiv:  0812.4477}.

\bibitem{ajt2} E. Andreini, Y. Jiang, H.-H. Tseng, ``Gromov-Witten theory
of product stacks,'' {\tt arXiv:  0905.2258}.

\bibitem{ajt3} E. Andreini, Y. Jiang, H.-H. Tseng, ``Gromov-Witten theory
of etale gerbes, i:  root gerbes,'' {\tt arXiv:  0907.2087}.

\bibitem{t1} H.-H. Tseng, ``On degree zero elliptic orbifold
Gromov-Witten invariants,'' Int. Math. Res. Not. IMRN 2011 2444-2468,
{\tt arXiv:  0912.3580}.

\bibitem{gtseng1} A. Gholampour, H.-H. Tseng, ``On Donaldson-Thomas invariants
of threefold stacks and gerbes,'' Proc. Amer. Math. Soc. {\bf 141} (2013) 
191-203, {\tt arXiv:  1001.0435}.

\bibitem{xt1} X. Tang, H.-H. Tseng, ``Duality theorems of \'etale gerbes
on orbifolds,'' Adv. Math. {\bf 250} (2014) 496-569,
{\tt arXiv:  1004.1376}.

\bibitem{ast} O. Aharony, N. Seiberg, Y. Tachikawa, ``Reading between the
lines of four-dimensional gauge theories,'' {\tt arXiv:  1305.0318}.

\bibitem{gmn} D. Gaiotto, G. Moore, A. Neitzke, ``Framed BPS states,''
{\tt arXiv:  1006.0146}.

\bibitem{yuji13} Y. Tachikawa, ``On the 6d origin of discrete additional data
of 4d gauge theories,'' {\tt arXiv:  1309.0697}.

\bibitem{hori94a} K. Hori, ``On global aspects of gauged Wess-Zumino-Witten
model,'' {\tt hep-th/9402019}.

\bibitem{hori94b} K. Hori, ``Global aspects of gauged Wess-Zumino-Witten
models,'' {\tt hep-th/9411134}. 

\bibitem{migdal1} A. Migdal, ``Recursion relations in gauge theories,''
Sov. Phys. JETP {\bf 42} 413-418,
(Zh. Eksp. Teor. Fiz. {\bf 69} (1975) 810-822).

\bibitem{rusakov1} B. Rusakov, ``Loop averages and partition functions in
$U(N)$ gauge theory on two-dimensional manifolds,''
Mod. Phys. Lett. {\bf A5} (1990) 693-703.

\bibitem{gt1} D. Gross, W. Taylor, ``Two-dimensional QCD is a string
theory,'' {\tt hep-th/9301068}.

\bibitem{cmr} S. Cordes, G. Moore, S. Ramgoolam, ``Lectures on 2d Yang-Mills
theory, equivariant cohomology, and topological field theories,''
{\tt hep-th/9411210}.

\bibitem{witten-2dgauge} E. Witten, ``On quantum gauge theories in two
dimensions,'' Comm. Math. Phys. {\bf 141} (1991) 153-209.

\bibitem{bt-ym} M. Blau, G. Thompson, ``Quantum Yang-Mills theory
on arbitrary surfaces,'' Int. J. Mod. Phys. {\bf A7} (1992) 3781-3806.

\bibitem{klimcik} C. Klincik, ``The formulae of Kontsevich and Verlinde from
the perspective of the Drinfeld double,''
Comm. Math. Phys. {\bf 217} (2001) 203-228,
{\tt hep-th/9911239}.

\bibitem{bt-2dgauge} M. Blau, G. Thompson, ``Lectures on 2d gauge
theories:  topological aspects and path integral techniques,''
{\tt hep-th/9310144}.

\bibitem{katz-qcd} E. Katz, G. Tavares, Y. Xu, ``Solving 2d QCD with an
adjoint fermion analytically,'' {\tt arXiv:  1308.4980}.

\bibitem{benini1} F. Benini, S. Cremonesi, ``Partition functions of
N=(2,2) gauge theories on $S^2$ and vortices,''
{\tt arXiv:  1206.2356}.

\bibitem{doroud1} N. Doroud, J. Gomis, B. Le Floch, S. Lee,
``Exact results in D=2 supersymmetric gauge theories,''
JHEP 1305 (2013) 093,
{\tt arXiv:  1206.2606}.

\bibitem{ggk} E. Gerchkovitz, J. Gomis, Z. Komargodski,
``Sphere partition functions and the Zamolodchikov metric,''
{\tt arXiv:  1405.7271}.

\bibitem{georgi} H. Georgi, {\it Lie algebras in particle physics},
second edition, Perseus Books, Reading, Massachusetts, 1999.

\bibitem{witten-dyons} E. Witten, ``Dyons of charge $e \theta / 2 \pi$,''
Phys. Lett. {\bf B86} (1979) 283-287.

\bibitem{vw} C. Vafa, E. Witten, ``A strong coupling test of S-duality,''
Nucl. Phys. {\bf B431} (1994) 3-77, {\tt hep-th/9408074}.

\bibitem{kaps} A. Kapustin, N. Seiberg, ``Coupling a QFT to a TQFT and
duality,'' {\tt arXiv:  1401.0740}.

\bibitem{dw1} R. Dijkgraaf, E. Witten, ``Topological gauge theories
and group cohomology,'' Comm. Math. Phys. {\bf 129} (1990) 393-429.

\bibitem{cdt} E. Sharpe, ``Analogues of discrete torsion for the M theory
three form,'' Phys. Rev. {\bf D68} (2003) 126004,
{\tt hep-th/0008170}.

\bibitem{dtrev} E. Sharpe, ``Recent developments in discrete torsion,''
Phys. Lett. {\bf B498} (2001) 104-110,
{\tt hep-th/0008191}.

\bibitem{aspinwall-dt} P. Aspinwall, ``A note on the equivalence of
Vafa's and Douglas's picture of discrete torsion,''
JHEP 0012 (2000) 029,
{\tt hep-th/0009045}.

\bibitem{stt} I. Shapiro, X. Tang, H.-H. Tseng, ``On the relative dual
of an $S^1$-gerbe over an orbifold,'' {\tt arXiv:  1312.7316}.

\bibitem{tsengpriv} X. Tang, H.-H. Tseng, private communication.

\bibitem{cmr-str} S. Cordes, G. Moore, S. Ramgoolam,
``Large N 2d Yang-Mills theory and topological string theory,''
Comm. Math. Phys. {\bf 185} (1997) 543-619,
{\tt hep-th/9402107}.

\bibitem{grry1} A. Gadde, L. Rastelli, S. Razamat, W. Yan,
``The 4d superconformal index from $q$-deformed 2d Yang-Mills,''
Phys. Rev. Lett. {\bf 106} (2011) 241602, {\tt arXiv:  1104.3850}.

\bibitem{grry2} A. Gadde, L. Rastelli, S. Razamat, W. Yan,
``Gauge theories and Macdonald polynomials,''
Comm. Math. Phys. {\bf 319} (2013) 147-193,
{\tt arXiv:  1110.3740}.

\bibitem{tonypriv} T. Pantev, private communication.





\end{thebibliography}
\end{document}